\documentclass[aps,prf,showpacs,floats,onecolumn,floats,superscriptaddress,floatfix]{revtex4}

\usepackage{graphicx}
\usepackage{bm}
\usepackage{amsfonts}
\usepackage{color}

\usepackage{amsmath}    
\usepackage{epsfig}
\usepackage{subfigure}  
\usepackage{hyperref}   
\usepackage{bm}
\usepackage{amssymb}

\newcommand{\be}{\begin{equation}}
\newcommand{\ee}{\end{equation}}

\newcommand{\rep}[1]{{\color{black}{#1}}}
\newcommand{\proof}[1]{{\color{black}{#1}}}

\def\p{\partial}

\newcommand{\ov}{\overline{ v}}
\newcommand{\ou}{\overline{ u}}

\newcommand{\obu}{\overline{\boldsymbol{u}}}
\newcommand{\oP}{\overline{{ p}}}

\newcommand{\oPi}{\overline{{\Pi}}^\Delta}

\newcommand{\Pileo}{{\overline{\Pi}}^\Delta_{L}}

\newcommand{\otau}{\overline{{\tau}}^\Delta}

\newcommand{\bk}{\boldsymbol{k}}
\newcommand{\bx}{\boldsymbol{x}}
\newcommand{\by}{\boldsymbol{y}}
\newcommand{\br}{\boldsymbol{r}}
\newcommand{\bu}{\boldsymbol{u}}
\newcommand{\bv}{\boldsymbol{v}}
\newcommand{\bhu}{\boldsymbol{\hat{u}}}

\newcommand{\bq}{\boldsymbol{q}}
\newcommand{\bp}{\boldsymbol{p}}

\newcommand{\rocaddress}{Department of Mechanical Engineering, University of Rochester, Rochester, New York, USA}

\newcommand{\romeaddress}{Department of Physics and INFN, University of 
Rome ``Tor Vergata'', via della Ricerca Scientifica 1, 00133, Rome, Italy.}

\begin{document}

\title{Energy transfer in turbulence under rotation \footnote{postprint version of the manuscript published in Phys. Rev. Fluids 3, 034802, (2018)}}

\author{Michele Buzzicotti} 
\email{michele.buzzicotti@roma2.infn.it}
\affiliation{\romeaddress}
\author{Hussein Aluie} 
\email{hussein@rochester.edu}
\affiliation{\rocaddress}
\author{Luca Biferale} 
\email{biferale@roma2.infn.it}
\affiliation{\romeaddress}
\author{Moritz Linkmann}
\email{linkmann@roma2.infn.it}
\affiliation{\romeaddress}

\begin{abstract}
It is known that rapidly rotating turbulent flows are characterized by the emergence of simultaneous upscale and downscale energy transfer. Indeed, both numerics and experiments show the formation of large-scale anisotropic vortices together with the development of small-scale dissipative structures. However the organization of  interactions  leading to this complex dynamics remains unclear.
Two different mechanisms are known to be able to transfer energy upscale in a turbulent flow. The first is characterized by two-dimensional interactions among triads lying on the \proof{two-dimensional, three-component (2D3C)}/slow manifold, namely on the Fourier plane perpendicular to the rotation axis. The second mechanism is three-dimensional and consists of interactions between triads with the same sign of helicity (homochiral).
Here, we present a detailed numerical study of rotating flows using a suite of high Reynolds number direct numerical simulations (DNS) within different parameter regimes to analyze both upscale and downscale cascade ranges. 
We find that the upscale cascade at wave numbers close to the forcing scale is generated by increasingly dominant homochiral interactions which couple the three-dimensional bulk and the 2D3C plane. This coupling produces an accumulation of energy in the 2D3C plane, which then transfers energy to smaller wave numbers thanks to the two-dimensional mechanism. In the forward cascade range, we find that the energy transfer is dominated by heterochiral triads and is dominated primarily by interaction within the fast manifold where $k_z\ne 0$.
We further analyze the energy transfer in different regions in the real-space domain. In particular, we distinguish high-strain from high-vorticity regions and we uncover that while the mean transfer is produced inside regions of strain, the rare but extreme events of energy transfer occur primarily inside the large-scale column vortices.
\end{abstract}

\pacs{}
\maketitle

\section{Introduction}

Understanding the role of rotation on the dynamics of turbulent flows is a challenging theoretical problem with a high relevance in several practical contexts, from natural (i.e. astrophysical and geophysical \cite{barnes2001, cho2008atmospheric}), to industrial flows \cite{dumitrescu2004}. When a fluid is under rotation, the Coriolis force
in the momentum equation competes with the nonlinear inertial term.
The strength of rotation is quantified by the Rossby number, $Ro$, defined as the ratio of the fluid time scale to the rotation time scale. In particular, for any fixed Reynolds number, $Re$, smaller Rossby numbers are correlated with more important rotation in the flow's evolution \cite{greenspan1968, davidson2013}. One of the most prominent effects of strong rotation is the formation of large-scale coherent axisymmetric vortical structures parallel to the rotation axis. This phenomenon is associated with a two dimensionalization of the flow with almost no fluctuations in the direction of the rotation axis \cite{biferale2016prx}.
In the limit of $Ro \rightarrow 0$, the nonlinear advection term in in the Navier-Stokes equations (NSE) can be neglected over short  time scales $t=O(Ro)$ (e.g. Ref. \cite{chen2005resonant}). From the resulting linear vorticity equation, it is then possible to explain the existence of the columnar $2$-dimensional structures, (Taylor-Proudman theorem, see Ref. \cite{davidson2015turbulence}). However, for any small $Ro$, the secular nonlinear dynamics becomes significant over long time scales $t=O(1)$ which requires a multiple-scale analysis \cite{greenspan1968,waleffe93} to account for resonant interactions. The description of the flow over yet longer times $t=O(1/Ro)$ requires higher order asymptotic theories that can be unwieldy \cite{chen2005resonant}. This limits our understanding of high $Re$ rotating flows, especially in the small $Ro$ limit over long time scales or in the moderate $Ro$ regime.
In particular, a thorough characterization of interactions between the quasi-2D turbulence and the 3D background is still lacking, which motivates our work here.

In addition to its theoretical appeal, the moderate $Ro$ regime is particularly important in many natural and industrial flows. For example, the Rossby number for the synoptic scales at mid-latitude of atmospheric and oceanic flows affected by the rotation of the Earth or for the plasma dynamics in convective zone of the Sun is $Ro \approx 0.1\div 1$ \cite{pedlosky2013, miesch2000three}. At the same time the Reynolds number in these systems is very large, and the flows are in a fully developed turbulent state.
High $Re$ turbulent flows with Rossby $O(1)$ have been extensively investigated in recent years using modern experimental \cite{lamriben2011direct, morize2005decaying, campagne2016turbulent, gallet2014scale} and numerical \cite{biferale2016prx, mininni2010rotating} techniques. Empirically, it is clear that eddy formation is greatly influenced by bulk rotation, with a tendency toward two dimensionalization and an increasing anisotropy in the flow, (see Refs. \cite{sagaut2008, godeferd2015} for a review), although many fundamental questions are still open. The nonlinear mechanism leading to such a state is still not well understood and we lack a theoretical prediction for the scaling law of anisotropic spectra \cite{smith1999transfer, clark2015spatio}. Note that the theoretical complexity of the problem is enhanced by the fact that the Coriolis force has an `indirect' influence on the energy transfer, since it does not even enter in the kinetic energy budget.

Rotation gives rise to inertial waves with a frequency, $\omega$, between zero and two times the rotation rate, $2\Omega$. In the rotating frame of reference, inertial waves appear as solutions to the momentum equation in the linearized regime.
Therefore, one can think to model rapidly rotating turbulence as superposition of inertial waves with a short period perturbed by the nonlinear interactions. According to this idea, the fluid velocity on a long time evolution is given by high frequency inertial waves modulated by the coupling with the slow frequencies of the large scales eddies. 
Although separating waves from eddies in a turbulent flow has been considered impossible in the past \cite{stewart1969turbulence}, significant progress has been made by applying resonant wave theory \cite{waleffe93, galtier2003weak}. It has been shown that the equations for the slow eddies contain as a subset the two-dimensional Navier-Stokes equations for the vertically averaged velocity fields \cite{babin1996, embid1996}. Recently it has been proved rigorously that \proof{the two-dimensional, three-component 2D3C} (vertically averaged and possibly turbulent) solutions are stable to vertically dependent perturbations ~\cite{gallet2015exact}. 
Using the `instability hypothesis' \cite{waleffe92} it has been argued that resonant triadic interactions should drive the flow to become quasi-two-dimensional \cite{waleffe93, smith1999transfer, clark2014quantification, alexakis2015rotating}. Moreover, the development of experimental techniques such as particle image velocimetry (PIV,~\cite{adrian1991particle}), allowed a space-time sampling of the velocity field giving the possibility of quantifying anisotropy and identifying inertial waves present in rotating turbulent flows \cite{campagne2015disentangling}.
However, resonant wave theories are only valid when the rotation period is much shorter than the eddy turnover time at all scales, hence their approximations break down at small scales for sufficiently large Reynolds number. 
The regime of validity of the resonant wave theory under varying Reynolds and Rossby numbers has been investigated in \cite{chen2005resonant}. \proof{One concludes that by decreasing Rossby number, at any fixed $Re$, the decoupling of the 2D3C modes from the fast-3D manifold becomes more and more efficient.}

Another open question about 3D turbulence under rotation is concerned with the direction of the energy cascade. 
In this context, evidence of a dual cascade of energy has been found in both numerical simulations \cite{smith1999transfer, sen2012anisotropy} and experiments~\cite{campagne2014direct}. Namely, 3D turbulent flows under rotation develop a non-zero energy flux going simultaneously upward and downward from the forcing scale.
In 2D turbulence, due to the existence of two quadratic invariants; energy and enstrophy (integrated squared vorticity), the presence of an upscale energy cascade has been demonstrated \cite{kraichnan1967} and anticipated \cite{fjortoft1953} for many years. In the 3D case, due to the absence of the second quadratic invariant (enstrophy), we do not have a theoretical prediction on the direction of the energy flux, which in principle can be directed toward the small scales, as confirmed by the results of two-point closure \cite{brissaud1973, andre1977influence} and direct numerical simulations \cite{chen2003joint, chen2003intermittency}.
Recently, however, there seems to be growing evidence that an upscale transfer mechanism in 3D does indeed exist. Indeed, an upscale energy flux has been observed at high Reynolds numbers in several flows, such as in geophysical flows subject to rotation \cite{smith1996crossover, mininni2009scale}, in shallow fluid layers \cite{nastrom1984kinetic,celani2010turbulence}, in realistic circulation in the North Atlantic Ocean \cite{Aluieetal17}, and in conducting fluids and plasmas \cite{alexakis2006inverse,mininni2007inverse}. Furthermore, it has been recently shown that in all 3D flows there is always a subset of nonlinear interactions coupling modes with the same chirality which transfer energy systematically toward the large scales \cite{biferale2012inverse,biferale2013split}.

The goal of this work is to gain a better understanding of the mechanisms leading the energy transfer in homogeneous-isotropic-incompressible turbulence subject to a uniform background rotation. We aim to asses whether the upscale flux is produced by purely two-dimensional interactions or by three-dimensional channels with definite chirality. To this end, we perform a Fourier space analysis of the spectra and fluxes using both the slow-fast \cite{waleffe93, smith1999transfer} and helical \cite{waleffe92, biferale2012inverse} decompositions to measure their relative weight on the total transfer. In the second part of this work, we  extend the analysis to  physical space. In particular we quantify how much of the mean energy transfer (as measured in Fourier space) comes from regions dominated by vorticity (i.e. inside the Taylor columns) or by strain. Note that while the mean value of the physical space flux (averaged over the whole domain) is analytically equivalent to the Fourier space flux, the spatially local values are not trivially related to the spectral flux in Fourier space.

In this paper, we analyze data from high-resolution direct numerical simulationson up to $2048^3$ collocation points. 
The dataset is composed of \rep{three different sets of simulations}, in the first set we used a small-scale forcing peaked at $k_f=40$ and a rotation rate $\Omega=50$, in the second set we used a large-scale forcing at $k_f=4$ and a rotation rate $\Omega=10$, \rep{and in the third set we force again at small-scale $k_f=40$ but increasing the rotation rate at $\Omega=100$}. In the first set of simulations we have the development of a backward energy transfer directed from the forcing to the largest scale of the system while in the second configuration we have a dual cascade of energy, which goes from the forcing scales toward both the smallest and largest scales of the system. \rep{The third set is used to benchmark the property of the backward energy transfer observed in the first set at changing of the Rossby number. In the first two cases the Rossby number is of the order $Ro \sim 0.1$, while in the third case the Rossby number is of the order of $Ro \sim 0.05$}.

\section{Numerical Simulations}
We perform direct numerical simulations (DNS) of Eqs. (\ref{eq:navierstokes}) governing an incompressible rotating fluid in a triply periodic domain of size $L=2\pi$ using a fully dealiased parallel 3D pseudospectral code using grids of up to $N^3 = 2048^3$ collocation points. The time integration has been performed with the second-order Adams-Bashforth scheme with viscous term integrated implicitly. 
The  governing equations for a fluid in the rotating frame can be written as
\begin{equation}
\label{eq:navierstokes}
\begin{cases}
\partial_t \bm{u} + \bm{u} \cdot \nabla \bm{u} +2{\bm \Omega} \times {\bm u}
= - \nabla p + \nu\Delta\bm{u}+ \alpha \Delta^{-1}\bm{u} + \bm{f} \\
\nabla \cdot \bm{u} = 0,
\end{cases}
\end{equation}
where $\nu$ is the kinematic viscosity, the term $2{\bm \Omega} \times {\bm u}$ is the Coriolis force produced by rotation, and $\bm \Omega = \Omega \hat{z}$ is the angular velocity with frequency $\Omega$ around the rotation axis $\hat{z}$. The fluid density is constant and absorbed into the definition of pressure $p$. The linear friction term $\alpha \Delta^{-1}\bu$, with hypo-viscosity coefficient $\alpha$, is used only to prevent the formation of a condensate at the lowest Fourier modes and it is projected onto the subset of wave numbers where $ |\bk|\le 2$.  \rep{The external forcing, $\bm{f}$, 
is a time-correlated forcing given by a second-order Ornstein-Uhlenbeck process, see Ref. \cite{sawford1991reynolds}}.
The centrifugal force, $-\bm \Omega \times \bm \Omega \times (\br - \br_0) = \nabla \left [ \frac{1}{2}	(\bm \Omega \times (\br - \br_0))^2 \right ]$, is absorbed into the pressure $p$ which enforces the incompressibility condition $\nabla \cdot \bm{u} = 0$. 

Rotation breaks the statistical isotropy of the flow above, but not its statistical homogeneity. \rep{Rotating flows give rise to inertial waves which can be seen by considering the inviscid ($\nu=0$) and linear [$(\bu \cdot \nabla \bu)\rightarrow 0$] limits of Eq.~\eqref{eq:navierstokes} written for the vorticity,

\begin{equation}
\partial_t (\nabla \times \bm {u}) = 2 \left ( {\bm{ \Omega} \cdot \nabla} \right ) {\bm u}.
\label{eq:linearizedRotatingNS}
\end{equation}
}
This equation has a general solution given by the superposition of inertial waves, of the form;

\begin{equation}
\bu(\bx,t) = \sum_{\bk}  {\bf h_\pm}(\bk) e^{i[\bk \cdot \bx - \omega_\pm(\bk) t ]}
\label{eq:helicalwavesolution}
\end{equation}
where ${\bf h}_{\pm}(\bm{k})$ are the orthogonal eigenmodes of the curl operator \cite{greenspan1968}, and the wave frequencies, $\omega_\pm$, are given by the dispersion relation,
\begin{equation}
\omega^{\pm}(\bk)=\pm 2 \Omega \frac{k_z}{|\bk|},
\label{eq:dispersrelation}
\end{equation}
where $k_z$ is the direction parallel to the rotation axis. 
There are two waves per wavevector with opposite sign of helicity, the right-handed waves propagating in the direction of $\bk$ and the left-handed waves propagating in the $-\bk$ direction. In rotating flows, the dynamics is regulated by two non-dimensional control parameters, the Reynolds and the Rossby numbers which can be written as,
$$Re = \frac{UL_f}{\nu} \qquad Ro = \frac{( \varepsilon_f k_f^2 )^{1/3}}{\Omega} \ ,$$
where $L_f\sim 1/k_f$ is the forcing scale, $U$ is the velocity at the forcing scale and $\varepsilon_f$ is the rate of energy input.
The Rossby number represents the ratio between the rotation time scale $\tau_{\Omega} = 1/\Omega$ and the flow time scale at the forcing scale, $(\varepsilon_f k_f^2)^{-1/3}$. In the limit of large Rossby numbers, $Ro\gg1$, the flow can evolve freely under its own internal dynamics without being influenced by rotation.
In the $Ro\lesssim 1$ regime, we can expect to observe effects of rotation on the dynamics
at wavenumbers satisfying $ (\varepsilon  k^2)^{-1/3}  \gtrsim \tau_{\Omega}$, 
where $\varepsilon$ is the dissipation rate. 
The largest wavenumber satisfying this condition is  called the Zeman scale, $k_{\Omega} = (\Omega^3/ \varepsilon)^{1/2}$.
Phenomenologically, it is known that when the Zeman wavenumber is larger than that of forcing, $k_f$, the flow develops an upscale energy cascade toward $k < k_f$ and a simultaneous forward cascade toward  small scales $k> k_f$ \cite{biferale2016prx, mininni2009scale}. Comparing the definition of the Rossby number and the Zeman scale, we can see that the condition $k_{\Omega} > k_f$ requires $Ro < 1$. Our numerical simulations are set in this regime and exhibit simultaneous energy transfer toward $k > k_f$ and $k < k_f$. We shall now discuss in more detail our numerical simulations and the parameter regimes they are in.

\subsection{Dataset description}
\label{sect:data}

We consider two sets of simulations, which we call A and B (see Table~\ref{tbl:simulations}). In both cases, the ratio between the Zeman scale and the forcing scale is $k_{\Omega}/ k_f \sim 10$ and $Ro \approx 0.1$. The difference between the two cases is the scale at which the flow is forced, which affects the cascade range (upscale or downscale) being captured in our simulation. Set A is forced at $k_f=40$ and exhibits an upscale energy transfer from $k_f$ to the smallest Fourier modes, while the forward cascade is quickly dissipated by viscosity due to the limited wavenumber range beyond $k_f$. Set B, on the other hand, is forced at $k_f=4$ and exhibits a clear simultaneous upscale and downscale energy transfer. \rep{The reason of having two sets of simulations with the forcing at different scales comes from the difficulty of performing a simulation able to resolve simultaneously both inverse and direct energy cascade with a well-extended range of scales. In the following we use set A to analyze the properties of the energy transfer in the inverse cascade regime while set B is used to assess the properties of the forward cascade regime. Moreover in this work we have used a third set of simulations, set C, to validate the property of the backward cascade at changing of the Rossby number, in this last case the ratio between the Zeman scale and the forcing scale is $k_{\Omega}/ k_f \sim 50$ and $Ro \approx 0.05$.}

\begin{figure*}[htp]
\centering
\includegraphics[scale=0.59]{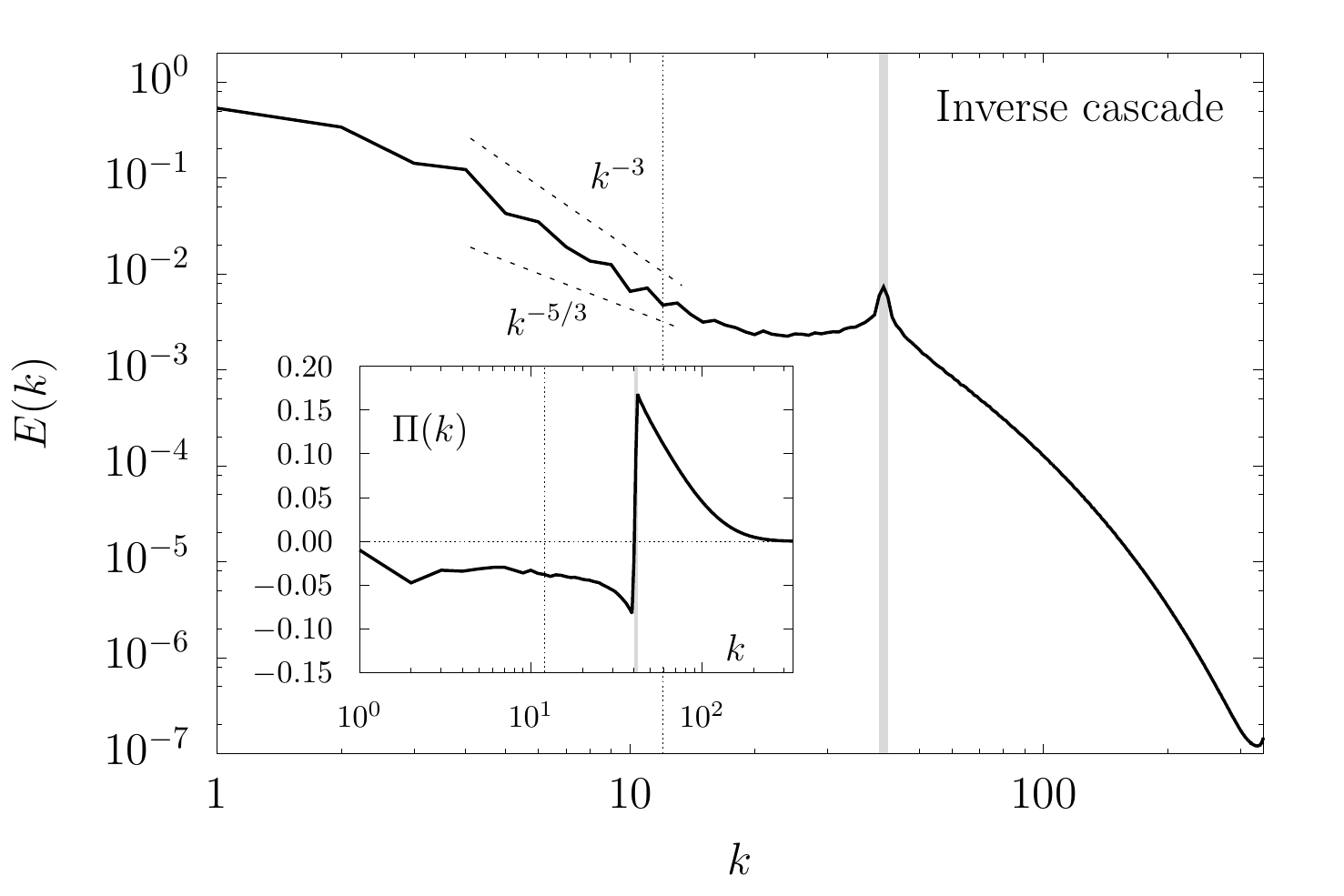}
\includegraphics[scale=0.59]{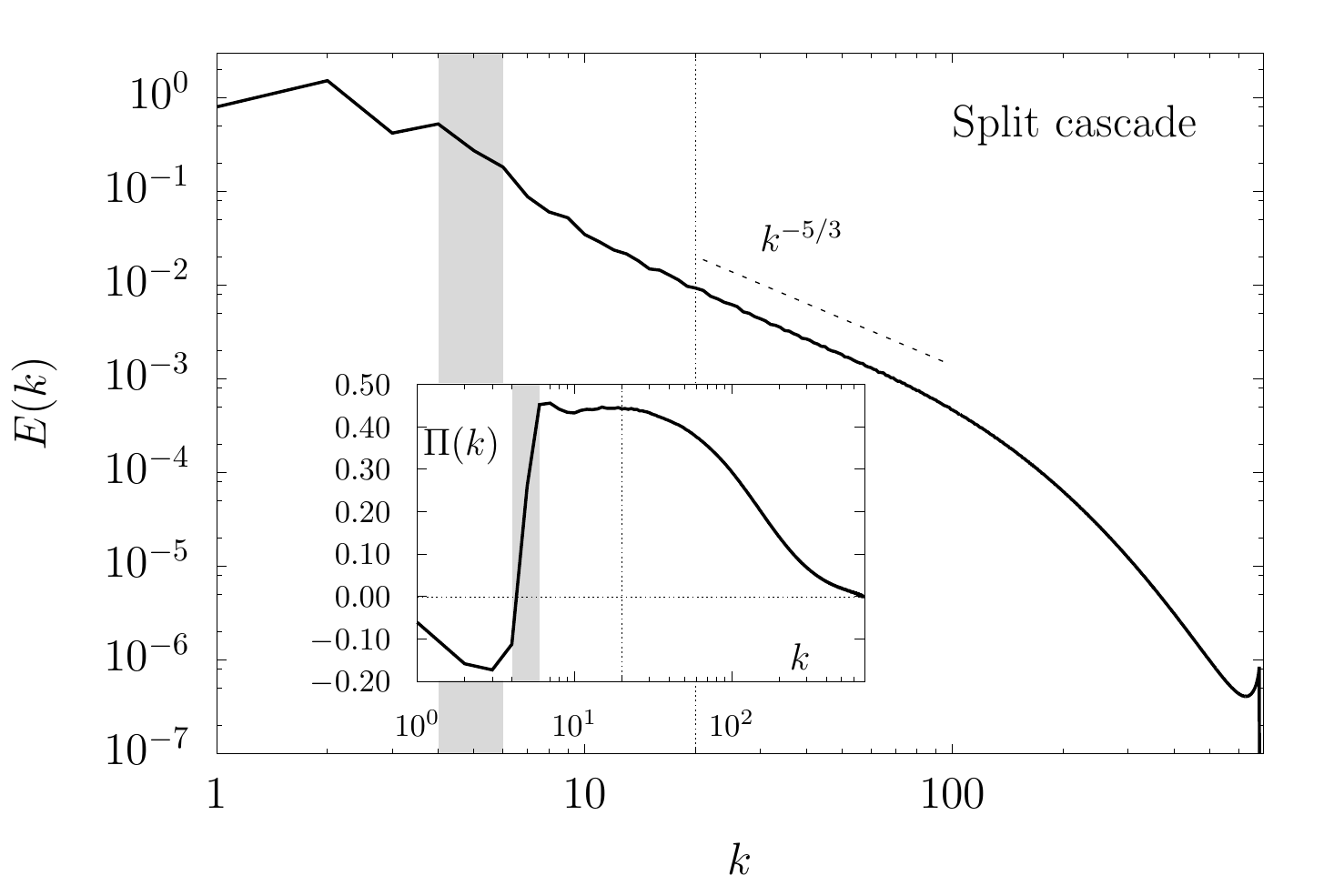}
\caption{Energy spectrum averaged in time in the stationary state for both simulations A (Inverse cascade) and B (Split cascade). In the insets, the energy fluxes averaged on the same time range are presented. The gray areas indicate the forced wavenumbers, \rep{while the dashed vertical lines represent the cutoff scale, $k_c$, used in the analysis of the \proof{sub-grid-scales (SGS)} energy transfer in both simulations (see Sec.~\ref{sec:sgs_transfer})}.}
\label{fig:sets_AB}
\end{figure*}

The energy spectra,
\be
\label{eq:spettro}
E(k)  = \frac{1}{2} \sum_{ k \le|\bk| < k+1} |\bhu(\bk)|^2 \ ,
\ee
 averaged in time after reaching a statistically steady state, are presented in Fig.~\ref{fig:sets_AB}.
In the upscale transfer regime of set A, the spectrum's slope seems to be closer to $\sim k^{-3}$ than to $\sim k^{-5/3}$.
This suggests that the dynamics over these scales cannot be described merely by 2D inverse cascade dynamics, and that there are important contributions from the energy injected into the 3D bulk (see Refs. \cite{smith1999transfer,sen2012anisotropy}). In contrast, the energy spectrum from dataset B in the downscale transfer regime has a slope close to $k^{-5/3}$, suggesting that the system is recovering isotropy at small scales. In the insets of the same figure \ref{fig:sets_AB} we also plot the energy fluxes,
\be
\label{eq:flusso}
\Pi(k)=-\sum_{|\bk|\le k}  ik_j \hat{u}^*_i(\bk)\sum_{\bp,\bq} \hat{u}_i(\bp)\hat{u}_j(\bq) \delta(\bp+\bq -\bk) \ ,
\ee
for both datasets, averaged over the same time period as the spectra.
The $\delta$ function constrains the nonlinear interactions to `triads' of wave-vectors that can form the sides of a triangle, $\bp+\bq-\bk=0$. The negative values of $\Pi(k)$ measured at scales $k<k_f$ indicate that in both sets A and B, there is an inverse energy cascade to large physical scales. In set B (split cascade), in addition to the upscale transfer, we also observe a constant positive flux  which indicates a forward cascade. This positive flux is quickly dissipated over $k>k_f$ in set A.

\begin{table*}
\begin{center}
\begin{tabular}{|cccccccccccccccc|}
Set & $N$ & $\Omega$ & $k_{f}$ & $\nu$ & $\varepsilon$ & $\varepsilon_f$& $ f_0$ & $\tau_f$ & $\eta/dx$& $\tau_\eta/dt$ & $Re_\lambda$ & $\lambda$ & $Ro$ & $T_0$ & $\alpha$\\
\hline
 (A: Inverse) & 1024 & 50 & [40:42] & $4 \times 10^{-4}$ & 0.25  & 0.275  & $8.8 \times 10^{-4}$ & 0.023 & 0.65 & 265 & 550 & 0.15 & 0.1 & 6.9 & 0.025\\
 (B: Split) & 2048 & 10 & [4:6]  & $1.5 \times 10^{-4}$ & 0.45  & 0.6  & 0.02 & 0.023 & 0.7 & 360 & 1500 & 0.13 & 0.1 & 3.0 & 0.1\\
{\rep{ (C: Inverse)}} & 1024 & 100 & [40:42] & $4 \times 10^{-4}$ & 0.135  & 0.15  & $8.8 \times 10^{-4}$ & 0.023 & 0.55 & 730 & 700 & 0.2 & 0.05 & 5.4 & 0.025\\
\end{tabular}
\end{center}
\caption{Eulerian dynamics parameters. 
  $N$: number of collocation points per spatial direction; 
  $\Omega$: rotation rate; 
  $k_{f}$: forced wavenumbers; $\nu$: kinematic viscosity; 
  $\varepsilon = \nu \int d^3x \sum_{ij} (\nabla_i u_j)^2$: viscous energy dissipation; 
  $\varepsilon_f = \int d^3x \sum_i f_i u_i$: energy injection;
  $f_0$: intensity of the Ornstein-Uhlenbeck forcing;
  $\tau_f$: decorrelation time of the forcing; 
  $\eta =(\nu^3/\varepsilon)^{1/4}$: Kolmogorov dissipative scale;
  $dx= L_0/N$: numerical grid spacing;
  $L_0= 2\pi$: box size;
  $\tau_\eta = (\nu/\varepsilon)^{1/2}$: Kolmogorov dissipative time;
  $dt$: Integration time step;
  $Re_{\lambda}=(u_0 \lambda)/\nu$: Reynolds number based on the Taylor micro-scale;
  $\lambda = (15 \nu u_0^2/\varepsilon)^{1/2}$: Taylor micro-scale;
  $Ro = (\varepsilon_f k_f^2)^{1/3}/\Omega$: Rossby number defined in terms of the energy injection properties;
  $T_0 =u_0/L_0$: Eulerian large-scale eddy turn over time; 
  $\alpha$: coefficient of the damping term $\alpha \Delta^{-1}{\bm u}$.}
\label{tbl:simulations}
\end{table*}

\section{Slow-fast decomposition}
\label{sec:slow-fast-dec}

The velocity field in a rotating turbulent flow can be analyzed as a multiple time scale problem \cite{greenspan1968}. A fast time scale is associated with inertial waves while a slow time scale can be associated with the turbulent eddies which modulate the waves' amplitude. Motivated by this analysis framework, we decompose the velocity as \cite{greenspan1968,waleffe93}

\begin{equation}
\bu(\bx,t) = \sum_{\bk} b_{\pm}(\bk, t) {\bf h_\pm}(\bk) e^{i[\bk \cdot \bx - \omega_\pm(\bk) t ]}.
\label{eq:waveseddiesdecomp}
\end{equation}
Note that the inertial waves' time scale is bounded from below by $\tau_{\omega} \ge 1/(2\Omega)$, while the time-scale of turbulent eddies is expected to scale as $\tau_{eddy} \sim (\varepsilon  k^2)^{-1/3}$. For this reason the large-physical-scale evolution is dominated by the inertial waves while the influence of the nonlinear dynamics becomes larger at smaller physical-scales. 
From the dispersion relation  in Eq.~\eqref{eq:dispersrelation} it is clear that all wavenumbers lying in the Fourier space plane, $\bk_{\perp}=(k_x,k_y,k_z=0)$, perpendicular to the rotation axis, do not give rise to inertial waves. This two-dimensional submanifold, where $\omega_\pm(\bk_{\perp}) = 0$, is generally called the `slow' manifold, while the rest of the 3D domain, where $\omega_\pm(\bk) \ne 0$ is called the `fast' manifold \cite{greenspan1969, embid1998low, smith1999transfer}. In the following, we refer to the slow modes as $\bk_{_S}=\bk_{\perp}$, and to the fast modes as $\bk_{_F} = (k_x,k_y,k_z\ne 0)$. 
Energy in a turbulent flow under rotation tends to be transferred from the fast toward the slow manifold ($k_z=0$) \cite{mininni2009scale}, this behavior was predicted in the context of the `wave resonance theory' by the \proof{Eddy Damped Quasi-Normal Markovian (EDQNM)} model of Cambon and Jacquin \cite{cambon1989spectral} and using the `instability assumption' by Waleffe \cite{waleffe93}. 
However, the problem with the wave turbulence theories is that they can only predict a net transfer toward small value of $k_z$ but not strictly zero. Those theories predict that energy transfer toward modes with $k_z = 0$ vanishes and the resulting dynamics for the slow modes is completely decoupled, hence purely two-dimensional \cite{waleffe93, clark2014quantification, alexakis2015rotating}.
In this work, using the slow-fast decomposition, we numerically assess the relative transfer inside the two manifolds as well as the importance of their mutual interactions.
Following \cite{bourouiba2008model}, we divide the total Fourier space volume into two disjoint 
subsets $V$ and $W$:

\begin{align}
\label{eq:s-f-man}
V &= \left \{ \bk \,| \, k_x, k_y \text{ and } k_z = 0  \right \} \\ \nonumber
W &= \left \{ \bk \,| \, k_x, k_y \text{ and } k_z \ne 0  \right \} \,
\end{align}
which correspond to the slow and the fast manifolds, respectively. 
Applying the same decomposition to the three dimensional velocity $\bhu(\bk)$ we can separate the field into the slow $\bhu_{_S}$ and the fast $\bhu_{_F}$ components, in the following way;

\be
\label{eq:s-f-decomp-u}
\bhu(\bk) = \begin{cases} \bhu_{_F} (\bk)  &\text{ if } \bk = \bk_{_F} \in W \\
\bhu_{_S} (\bk)  &\text{ if }  \bk = \bk_{_S} \in V \ .
\end{cases}
\ee
The slow manifold consist of a two-dimensional, three-component (2D3C) field which evolves on the $\bk_{\perp}$ plane perpendicular to the rotation axis. The 2D3C field can be further decomposed into a pure 2D part and in its third passive component,

\be
\label{eq:2d3c-decomp-u}
\bhu_{_S}(\bk_{_S}) = \bhu_{2D}(\bk_{_S}) + \theta(\bk_{_S})  
\ee
where,

\be
\label{eq:2d3c-components}
\bhu_{_{2D}}(\bk_{_S}) = \begin{pmatrix} \hat{u}_x(\bk_{_S}) \\ \hat{u}_y(\bk_{_S}) \\ 0   \end{pmatrix}
\qquad \theta (\bk_{_S})=  \begin{pmatrix} 0  \\ 0  \\ \hat{u}_z(\bk_{_S}) \end{pmatrix} \ .
\ee
Assuming a complete decoupling between slow and fast dynamics, the physical space 2D3C-Navier-Stokes equations for the slow incompressible flow read;

\begin{align}
\label{eq:2d3c-nse}
\begin{cases}
\partial_t \bu_{_{2D}} &= -(\bu_{_{2D}} \cdot \nabla) \bu_{_{2D}} - \nabla p + \nu \Delta \bu_{_{2D}} \ ,\\  
\partial_t \theta  &= -(\bu_{_{2D}} \cdot \nabla) \theta + \nu \Delta \theta  \ , 
\end{cases}
\end{align}
where $\bu_{_{2D}}$, $\theta$ are respectively the inverse Fourier transform of the fields in Eq.~\eqref{eq:2d3c-components} and $\nabla \cdot \bu_{_{2D}} = 0$. However a complete decoupling is only attained asymptotically in the limit of $Ro \rightarrow 0$ \cite{chen2005resonant}. In the set-up of  simulations A and B, we need to consider the nonlinear interactions between the two submanifolds, which can be written explicitly by projecting the full nonlinear term of the Navier-Stokes equations on the two different manifolds;

\begin{align}
\label{eq:slow-ns}
\partial_t \bu_{_S} + {\text P_{_S}}  \left (\nabla p \right) &= -\bu_{_S} \cdot \nabla \bu_{_S} - {\text P_{_S}} \left ( \bu_{_F} \cdot \nabla \bu_{_F} \right ) \\ 
\partial_t \bu_{_F} + {\text P_{_F}}  \left (\nabla p \right) &= -\bu_{_F} \cdot \nabla \bu_{_S} - \bu_{_S} \cdot \nabla \bu_{_F}- {\text P_{_F}} \left ( \bu_{_F} \cdot \nabla \bu_{_F} \right )
\label{eq:fast-ns}
\end{align}
where $\bu(\bx) = \bu_{_S}(\bx) + \bu_{_F}(\bx)$, $\text P_{_S}$ and $\text P_{_F}$ are respectively the projectors on the Fourier subsets V and W defined in Eq.~\eqref{eq:s-f-man}. Note that the two fields, $\bu_{_S}(\bx)$ and $\bu_{_F}(\bx)$, are separately divergence-free. 
Multiplying Eq.~\eqref{eq:slow-ns} by $\bu_{_S}$ and Eq.~\eqref{eq:fast-ns} by $\bu_{_F}$ 
we end up with an explicit formulation for the pointwise energy transfer in the physical space volume;

\begin{align}
\label{eq:slow-energy}
\partial_t \frac{1}{2} \bu^2_{_S} + \bu_{_S} {\text P_{_S}} (\nabla p) &= - \bu_{_S} \cdot \left (\bu_{_S} \cdot \nabla \bu_{_S} \right )  - \bu_{_S} \cdot ( {\text P_{_S}} \left [\bu_{_F} \cdot \nabla \bu_{_F} \right ]) \\
\partial_t \frac{1}{2} \bu^2_{_F} + \bu_{_F} {\text P_{_F}} (\nabla p) &= -\bu_{_F} \cdot \left (\bu_{_F} \cdot \nabla \bu_{_S} \right ) - \bu_{_F} \cdot \left (\bu_{_S} \cdot \nabla \bu_{_F} \right ) - \bu_{_F} \cdot \left ( {\text P_{_F}} [\bu_{_F} \cdot \nabla \bu_{_F}] \right ),
\label{eq:fast-energy}
\end{align}
note that on average $ \langle \bu_{_S} \cdot \bu_{_F} \rangle = 0$, hence the total energy is given by the sum of the energy contained in the sets V and W separately. 
It is important to stress that the three terms coupling $\bu_{_S}$ and $\bu_{_F}$ dynamically in eqs.~\eqref{eq:slow-energy}-\eqref{eq:fast-energy} are the only possible combinations between slow and fast modes, considering that the coupling can happen only amongst wavevectors satisfying the triadic condition (we cannot have a closed triad composed of two slow modes and a fast mode). Moreover, amongst the three coupling terms, only $\bu_{_F} \cdot \left (\bu_{_S} \cdot \nabla \bu_{_F} \right )$ is a real flux which conserves total energy when averaged over the total volume, while only the sum of the remaining two terms can be written as a total gradient and conserves energy over the entire volume:
\be
\langle\bu_{_F} \cdot \left (\bu_{_F} \cdot \nabla \bu_{_S} \right ) \rangle + \langle \bu_{_S} \cdot \left (\bu_{_F} \cdot \nabla \bu_{_F} \right ) \rangle  
= \langle \nabla \cdot \left ( \bu_{_S} \cdot  (\bu_{_F} \otimes \bu_{_F}) \right )\rangle = 0 \ .
\ee

\begin{figure*}
\hspace{-0.2cm}
\includegraphics[scale=0.45]{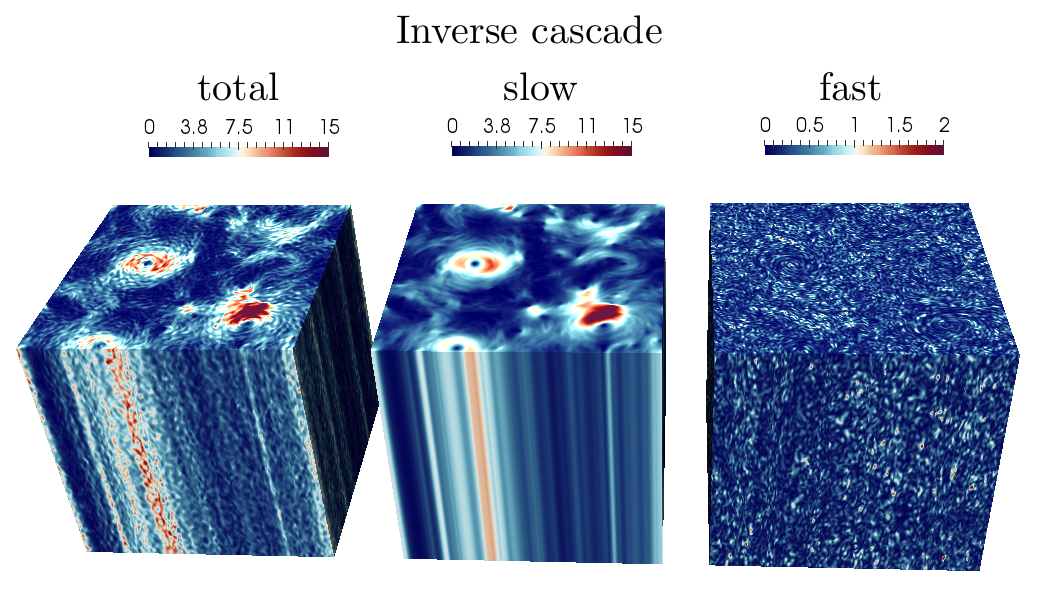}
\caption{Snapshot of module squared, $u^2(\bx)=|\bu(\bx)|^2$, of a total velocity field (left) and velocity decomposed on the slow (center) and fast (right) manifolds. 
Data correspond to simulation A.}
\label{fig:vel_1024}
\end{figure*}

\begin{figure*}
\hspace{-0.cm}
\includegraphics[scale=0.52]{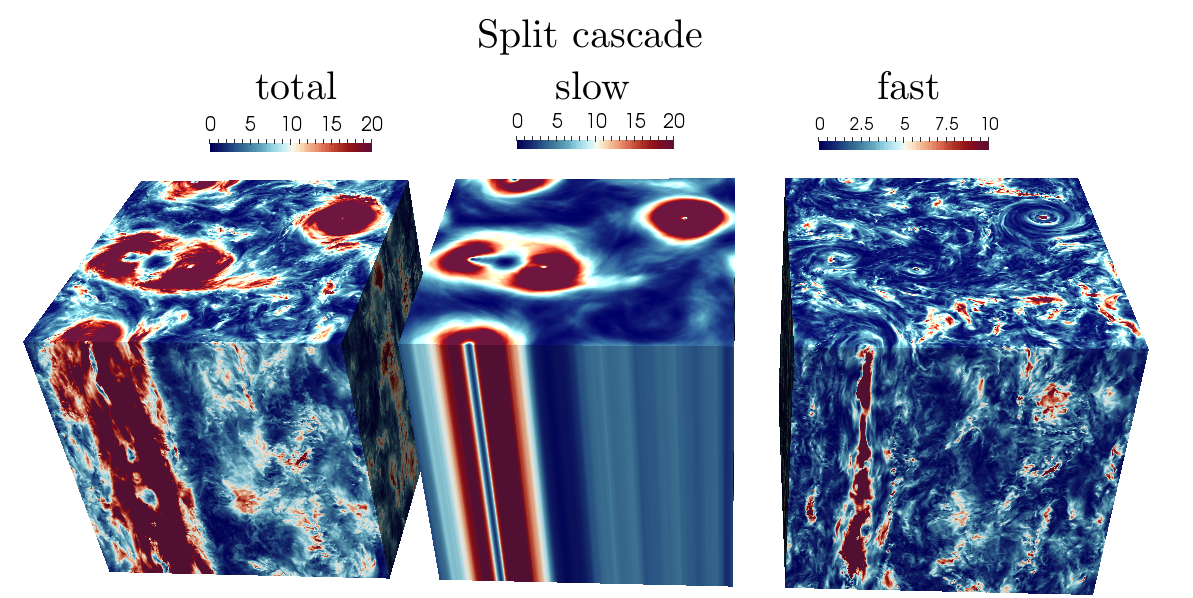}
\caption{Snapshot of module squared, $u^2(\bx)=|\bu(\bx)|^2$, of a total velocity field (left) and velocity decomposed on the slow (center) and fast (right) manifolds. 
Data correspond to simulation B.}
\label{fig:vel_2048}
\end{figure*}

\begin{figure*}
\centering
\includegraphics[scale=0.48]{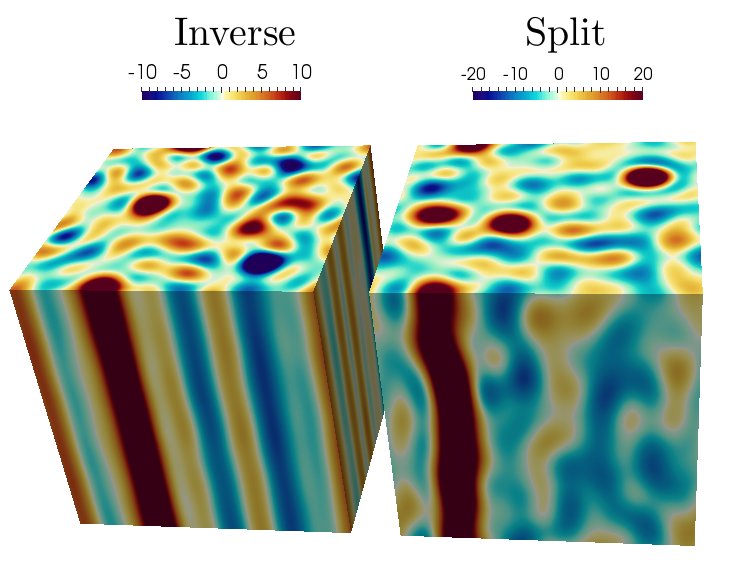}
\caption{Snapshot the vorticity along the direction of the rotation axis $z$, $\omega_z$, for the two sets of simulations, A (inverse) and B (split). The large-scale columnar structures are parallel to the direction of the external angular velocity $\bm \Omega = \Omega \hat{z}$. In particular the cyclonic structures, columns with positive (red) vorticity, rotate in the direction of $\bm\Omega$, while the negative (blue) columns observed in dataset A (inverse) are the anticyclonic vortices rotating in the inverse direction of $\bm\Omega$. \proof{The two vorticity fields are filtered using a sharp cutoff in Fourier space at scales $k_c=12$ and $k_c=20$ respectively for the inverse and the split cascade (see sect.\ref{sec:sgs_transfer}).}}
\label{fig:vort_AB}
\end{figure*}

\noindent
Going back to the Fourier space definition of the slow-fast modes, it is straightforward to notice that the two projected velocity fields are orthogonal: 
$\bhu_{_S}(\bk) \cdot \bhu_{_F}(\bk) = 0$. From this property we can rewrite the energy spectrum as the sum of three different components;

\be
E(k) = E_{_F}(k) + E_{_S}(k) = E_{_F}(k) + E_{2D}(k) + E_{\theta}(k) 
\label{eq:spectra2d3c}
\ee
where in the first step we used the decomposition in slow-fast modes,

\be
E_{_F}(k) = \frac{1}{2} \sum_{\substack{\bk \in W \\  k \le|\bk| < k+1}} |\bhu_{_F}(\bk)|^2 \, , \,\, E_{_S}(k) = \frac{1}{2} \sum_{\substack{\bk \in V \\  k \le|\bk| < k+1}} |\bhu_{_S}(\bk)|^2 \ ,
\ee
while in the second step we further decomposed the slow manifold in the purely 2D field and its $z$-component, see Eq.~\eqref{eq:2d3c-components};

\be
E_{2D}(k) = \frac{1}{2} \sum_{\substack{\bk \in V \\  k \le|\bk| < k+1}} |\bhu_{2D}(\bk)|^2 \, , \,\, E_{\theta}(k) = \frac{1}{2} \sum_{\substack{\bk \in V \\  k \le|\bk| < k+1}} |\theta(\bk)|^2 \ .
\ee

Fourier transforming all nonlinear terms in equations ~\eqref{eq:slow-energy}-\eqref{eq:fast-energy}, we end up with the different energy fluxes due to the different kind of interactions;

\be
\sum_{k'=1}^{k} \partial_t E(k',t) = \Pi_{_{S \leftrightarrows S}}(k,t) + \Pi_{_{F \leftrightarrows F}}(k,t) + \Pi_{_{F \leftrightarrows S}}(k,t) \ ,
\ee
where $\Pi_{_{S \leftrightarrows S}}$, $\Pi_{_{F \leftrightarrows F}}$ are energy fluxes which transfer energy only amongst modes which live respectively in the slow or in the fast manifold, while the term $\Pi_{_{F \leftrightarrows S}}$ represents the energy flux due to the coupling between slow and fast modes.

It is worth noticing that while $\Pi_{_{S \leftrightarrows S}}$ and $\Pi_{_{F \leftrightarrows F}}$ contain a single term, hence a single class of triadic interactions, namely;
 
\be
\Pi_{_{S \leftrightarrows S}}(k) = - \sum_{\substack{\bk \in V \\ |\bk|\le k}}  ik_j \hat{u}^*_i(\bk)\sum_{\bp,\bq \in V}  \hat{u}_i(\bp)\hat{u}_j(\bq) \delta(\bp+\bq -\bk) ,
\label{eq:fluxslowslow}
\ee  
and
\be
\Pi_{_{F \leftrightarrows F}}(k) = - \sum_{\substack{\bk \in W \\ |\bk|\le k}}  ik_j \hat{u}^*_i(\bk)\sum_{\bp,\bq \in W}  \hat{u}_i(\bp)\hat{u}_j(\bq) \delta(\bp+\bq -\bk) ,
\label{eq:fluxfastfast}
\ee  
the coupling interactions can be decomposed into three different contributions coming from the three different coupling terms;
\begin{align}
\nonumber
\Pi&_{_{F \leftrightarrows S}}(k) = -\sum_{\substack{\bk \in W \\ |\bk| \le k}}  ik_j \hat{u}^*_i(\bk)\sum_{\substack{\bp\in W \\ \bq\in V}}  \hat{u}_i(\bp)\hat{u}_j(\bq) \delta(\bp+\bq -\bk) \\ \nonumber
&- \sum_{\substack{\bk \in W \\ |\bk|\le k}}  ik_j \hat{u}^*_i(\bk)\sum_{\substack{\bp\in V \\ \bq\in W}}  \hat{u}_i(\bp)\hat{u}_j(\bq) \delta(\bp+\bq -\bk) \\ 
&- \sum_{\substack{\bk \in V \\ |\bk|\le k}}  ik_j \hat{u}^*_i(\bk)\sum_{\substack{\bp\in W \\ \bq\in W}}  \hat{u}_i(\bp)\hat{u}_j(\bq) \delta(\bp+\bq -\bk) \,.
\label{eq:fluxfastslow}
\end{align}

In Fig.~\ref{fig:vel_1024} and in Fig.~\ref{fig:vel_2048} a visualization of a real-space velocity field decomposed on the slow and fast manifolds is presented for both datasets. It is interesting to notice that in the split cascade simulation (dataset A, Fig.~\ref{fig:vel_2048}) we observe both the formation of 3D structures in the fast manifold inside the cores of the rotating vortexes as well as the big Taylor columns. While in the backward cascade regime (dataset B, Fig.~\ref{fig:vel_1024}) we observe only the formation of large-scale vortexes meaning that the energy cascading forward is quickly dissipated preventing the formation of smaller-scale structures. The vorticity along the direction of the rotation axis, $\omega_z(\bx)$ is shown in Fig.~\ref{fig:vort_AB}. From the latter we can distinguish cyclonic and anticyclonic vortices for both simulations A and B. In particular we can notice the presence of both cyclonic and anticyclonic structures in set A, while set B is dominated by big cyclonic vortexes. 
\rep{To highlight the formation of the large-scales vortices the two fields shown in Fig.~\ref{fig:vort_AB} have
been filtered using a sharp cutoff in Fourier space at $k_c=12$ for the inverse cascade and at $k_c=20$ for the split cascade (see sect.\ref{sec:sgs_transfer}).}

\begin{figure*}
\includegraphics[scale=0.59]{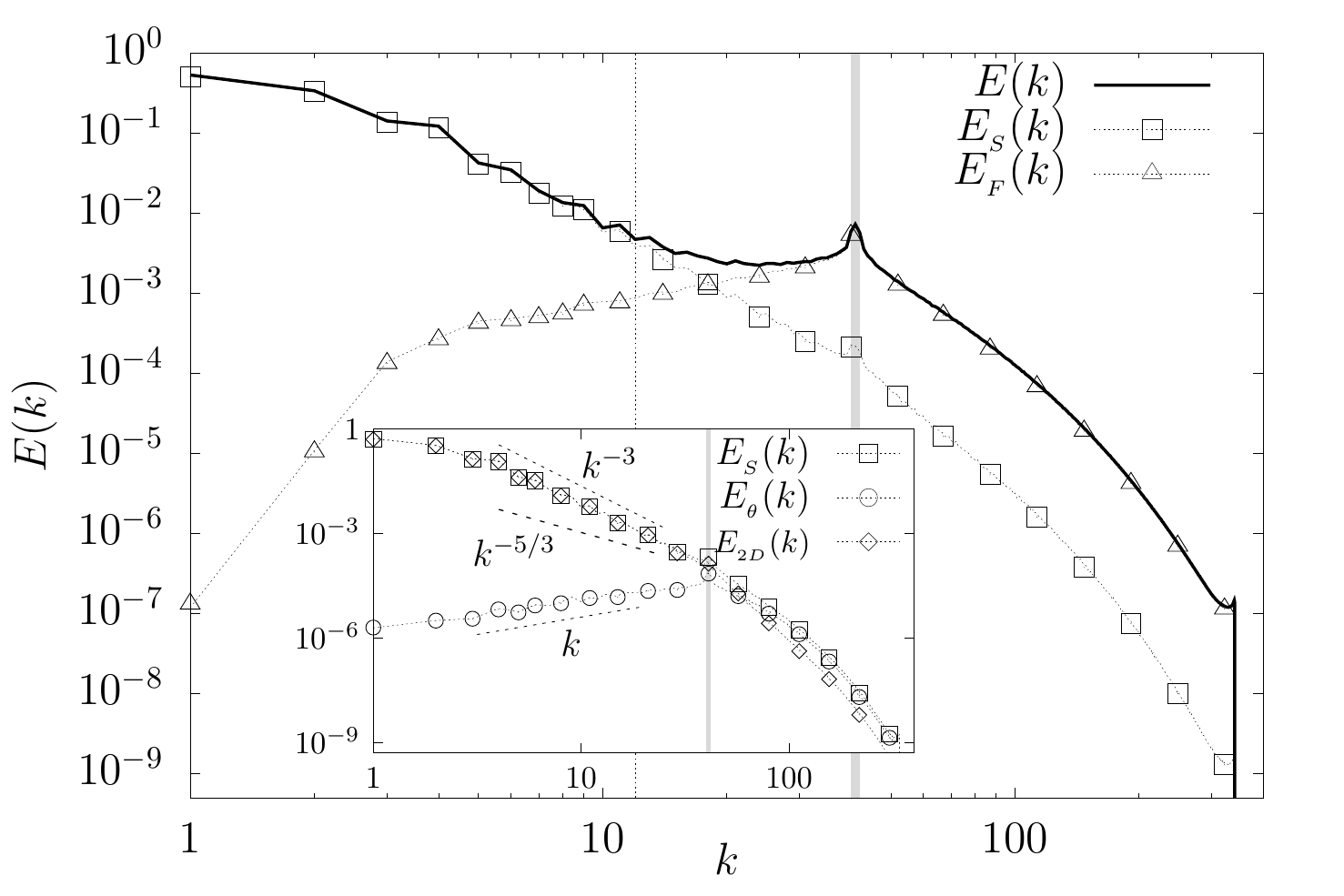}
\includegraphics[scale=0.59]{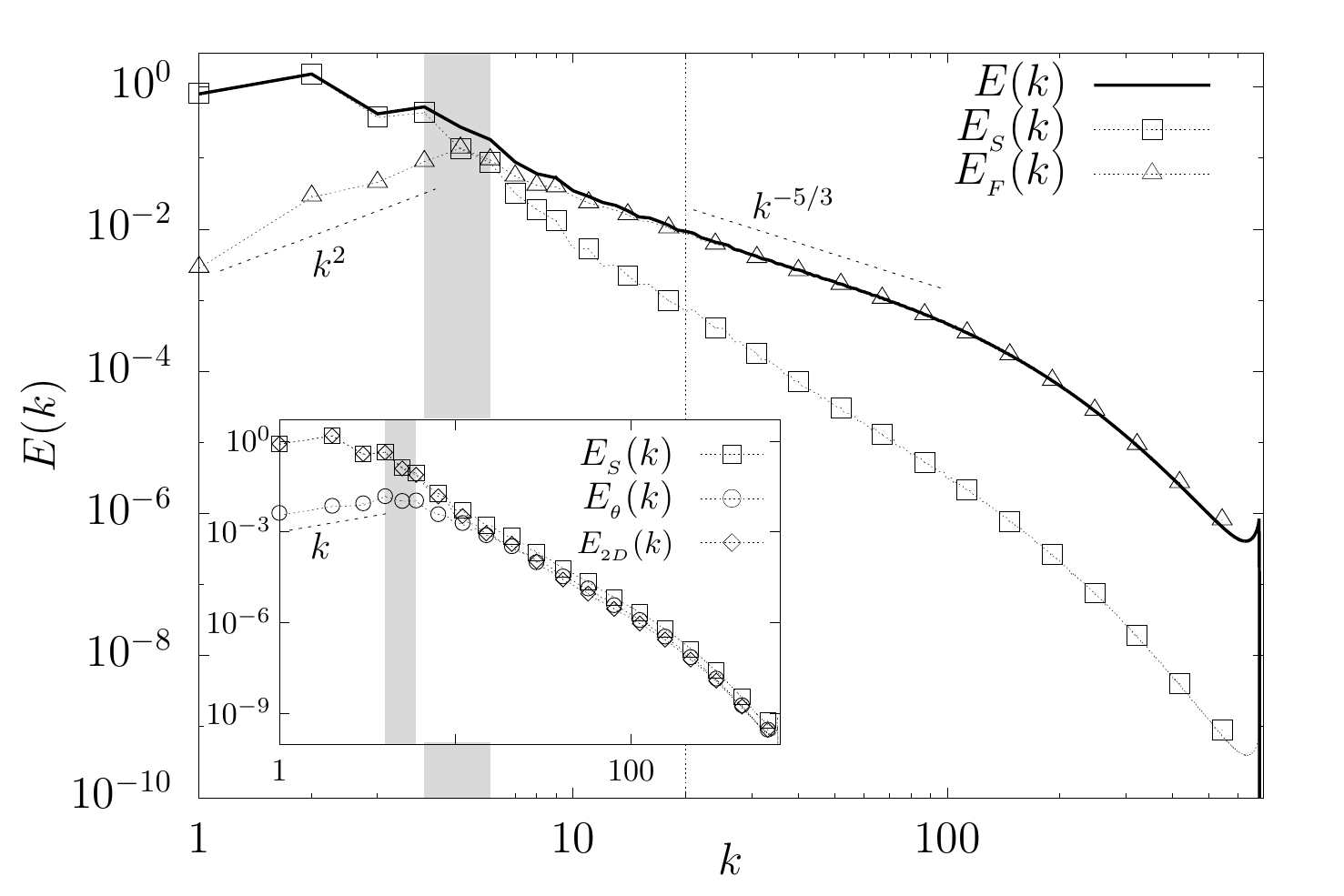}
\caption{Spectrum decomposition into fast and slow manifolds for set A (left panel) and set B (right panel). The gray areas represent the forced wavenumbers, \rep{the dashed vertical lines represent the cutoff scale, $k_c$, used in the analysis of the SGS energy transfer in both simulations (see Sec.~\ref{sec:sgs_transfer})}. In the insets there are the slow spectra further decomposed into their parallel $E_{2D}(k)$ and passive, $E_{\theta}(k)$, components.}
\label{fig:spectra}
\end{figure*}

\section{Helical decomposition}
\label{sec:hel-dec}

In this section we exploit the decomposition of any incompressible 3D flow into helical modes proposed in \cite{waleffe92,constantin1988}. This further decomposition allows us to clarify the analysis of the mechanism responsible of the energy transfer in a fluid under rotation. 
\rep{Decomposing the inverse flux in terms of the homochiral and heterochiral components is important to highlight the existence of a purely 3D mechanism contributing to the inverse cascade. It is indeed known that homochiral triads are always responsible of opening a channel with inverse energy cascade, even in homogeneous and isotropic turbulence \cite{alexakis2017helically,sahoo2018energy}. In the following we show that the same mechanism is at work also in rotating turbulence, at least in a range of scales not too far from the forcing scale. On the other hand when the inverse cascade is dominated by two dimensional mechanism, it is not sensitive to the sign of helicity \cite{biferale2017} and we measure the same contribution coming from hetero and homochiral interactions.}
Since $\bu(\bx)$ is a solenoidal vector field, its Fourier modes $\bhu(\bk)$ have only  two degrees of freedom, and we have 
\be
\bhu_{\bk}(t) = \bhu_{\bk}^+(t) + \bhu_{\bk}^-(t) = \hat u^+_{\bk}(t) {\bf h_+}(\bk) + \hat u^-_{\bk}(t) {\bf h_-}(\bk) \ ,
\ee
where ${\bf h_\pm}(\bk)$ are the normalized eigenvectors of the curl operator in Fourier space introduced in Eq.~\eqref{eq:helicalwavesolution}, (see  \cite{waleffe92}).
The helical decomposition thus decomposes the Fourier modes of the velocity field into two components, each of which satisfies    
\be
i\bk \times \bhu_{\bk}^{s_k}= s_kk \bhu_{\bk}^{s_k} \ ,
\ee
with $s_k = \pm$. The corresponding homo- and heterochiral energy fluxes are;
\begin{align}
\Pi^{\rm HO}(k)  &=  - \sum_{|{\bk}| \le k} \, \sum_{s\in\{+,-\}}
\hspace{-1em} 
\hat \bu^{s*}_{\bk} \, \cdot \, \hspace{-1em} \sum_{\bq=\bk-\bp}\hspace{-0.5em} 
 (i\bk \cdot \hat \bu^s_{\bp}) \hat\bu^s_{\bq} \ ,
\end{align}
\be
\Pi^{\rm HE}(k) = \Pi(k)-\Pi^{\rm HO}(k) \ ,
\label{eq:flux_helical_dec}
\ee
respectively, where $\Pi(k)$ is the total energy flux defined in Eq.~\eqref{eq:flusso}.

\section{Results: Fourier space analysis}
\label{sect:Fspace}

In this section we present the analysis of the Fourier space energy fluxes and spectra for both datasets following the slow-fast and the helical decomposition introduced in the previous sections. 

\subsection{Slow-Fast mode analysis}
In Fig.~\ref{fig:spectra} the total energy spectrum is split into the five different contributions coming from the slow-fast decomposition, see Eq.~\eqref{eq:spectra2d3c}. The left and right panels of Fig.~\ref{fig:spectra} are respectively obtained from the data of simulation set A and B, in both of them it is clear that the large-scale energy, i.e. $E(k)$ at $k< k_f$, is accumulated in the slow manifold, namely $E_{_S}(k)$, as predicted by \cite{cambon1989spectral, waleffe93} and observed in \cite{mininni2009scale}. Instead the energy in the forward cascade, hence $E(k)$ at $k>k_f$, is accumulated in the fast modes, $E_{_F}(k)$. 

The slow modes dominate the system at small wave numbers in the both the inverse and the split cascade regimes, while the fast interactions lead the forward cascade and contain most of the energy at large wavenumbers, that is, at small physical scales. In the insets of both panels in Fig.~\ref{fig:spectra} the decomposition of the slow modes into the pure 2D and passive component is shown. From this further decomposition we can see that the energy on the plane perpendicular to the rotation axis $\bk_{\perp}$ is distributed at small wavenumbers on the two-dimensional components of the velocity field lying on the plane, (i.e. $\hat{u}_{2D}(\bk_{_S})$), while the passive third component, $\theta(\bk_{_S})$, is involved only in the forward dynamics showing 
a 2D equipartition spectrum, $\sim k$, at small wavenumbers.  
This analysis of the slow manifold shows a behavior very similar to the one observed in the 2D3C Navier-Stokes system \cite{biferale2017}. However, as we can see from the inset of Fig.~\ref{fig:spectra}(left), the spectrum $E_{_S}(k)$ in the backward regime deviates from the $k^{-5/3}$ of the pure 2D3C system showing a $k^{-3}$ slope, see Refs. \cite{biferale2017, smith1999transfer}, meaning that the contribution of the 3D interactions is important for the dynamics of the backward cascade.

\begin{figure*}
\includegraphics[scale=0.59]{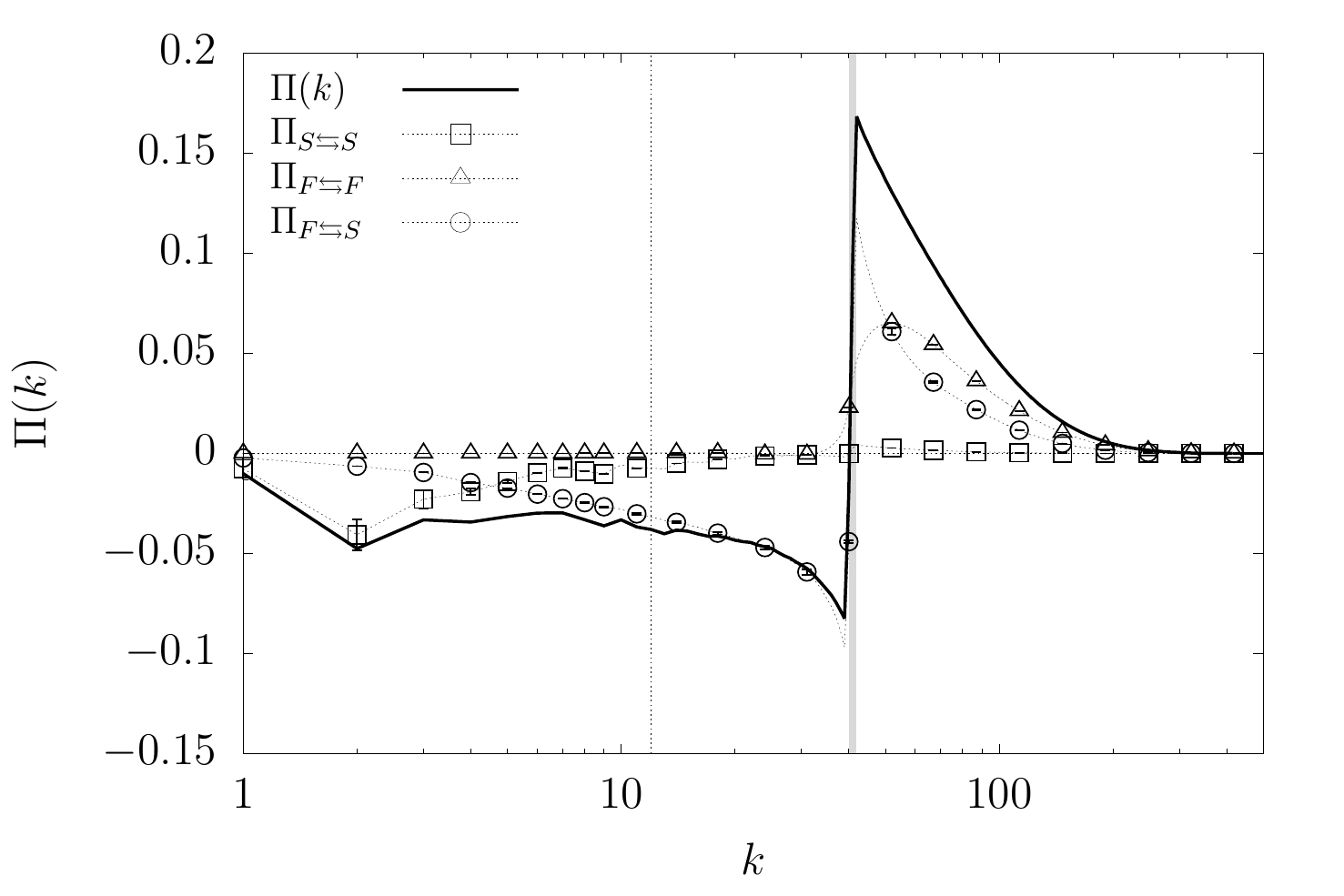}
\includegraphics[scale=0.59]{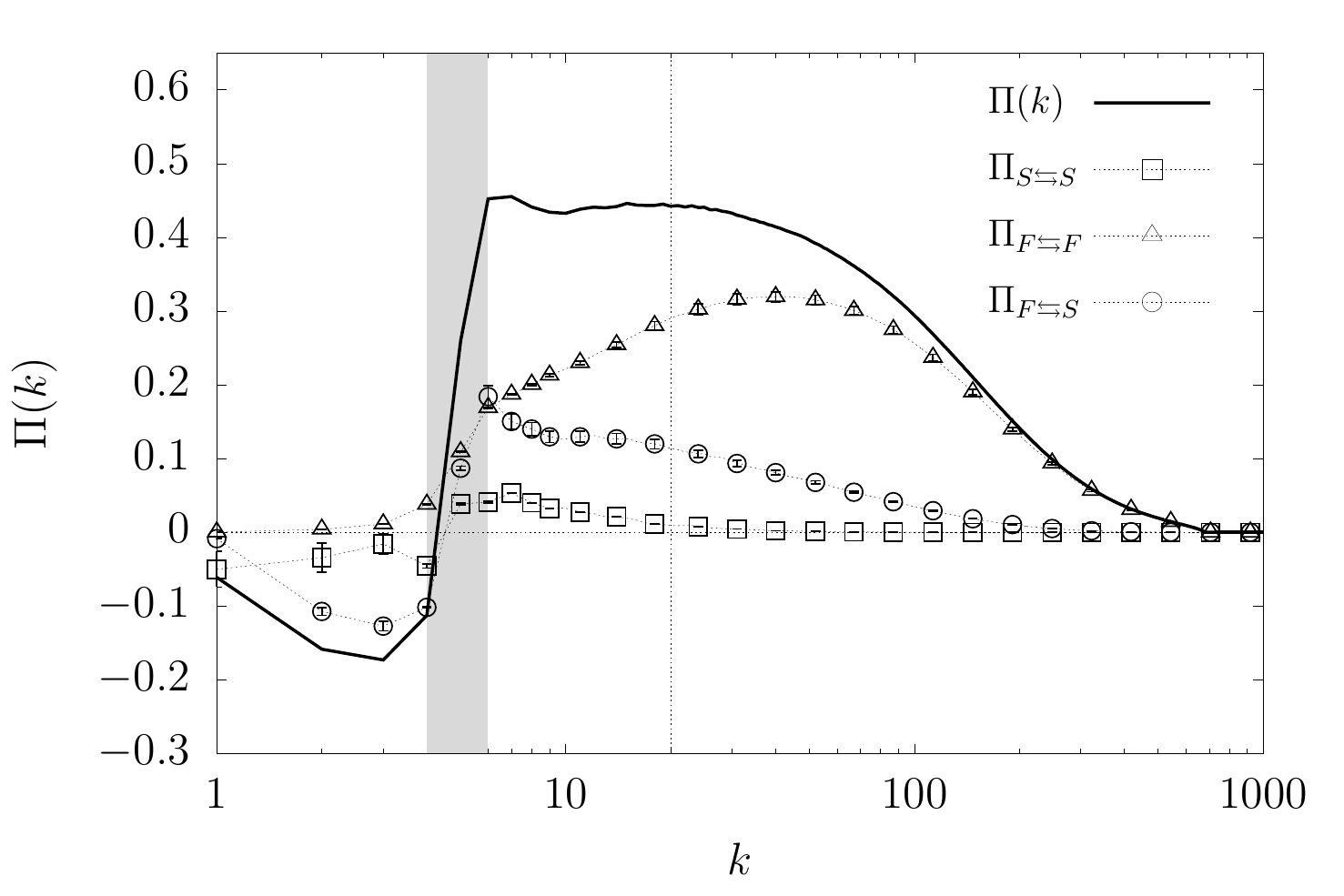}

\caption{Total energy flux $\Pi(k)$ (solid line) and fluxes decomposed on the different slow/fast interactions, namely slow-slow interactions, $\Pi_{_{S \leftrightarrows S}}(k)$ (empty squares), fast-fast interactions $\Pi_{_{F \leftrightarrows F}}(k)$ (empty triangles), coupling slow-fast interactions $\Pi_{_{F \leftrightarrows S}}(k)$ (empty circles). (Left panel) Data from simulation A, (right panel) data from simulation B. \rep{The dashed vertical lines represent the cutoff scale, $k_c$, used in the analysis of the SGS energy transfer (see Sec.~\ref{sec:sgs_transfer}), which are respectively in the inverse and in the direct cascade regimes for simulation set A and B}.}
\label{fig:flux_slow_fast}
\end{figure*}

In Fig.~\ref{fig:flux_slow_fast} the analysis of the decomposed energy fluxes is presented. Here we compare the total flux, $\Pi(k)$, with the different contributions coming from the slow-fast decomposition, namely; interactions defined inside the slow manifold Eq.~\eqref{eq:fluxslowslow}, inside the fast manifold Eq.~\eqref{eq:fluxfastfast} and the coupling ones Eq.~\eqref{eq:fluxfastslow}. From the flux decomposition we can see that the energy transfer from the forcing to the small wavenumbers is achieved thanks to two different processes, first the energy is transferred via coupling interactions between fast and slow modes, $\Pi_{_{F \leftrightarrows S}}^b(k)$ and only at very small wavenumbers is transferred by interactions on the slow manifold, $\Pi_{_{S \leftrightarrows S}}(k)$. These results suggest that in a turbulent flow under rotation, $Re\gg1$ and $Ro = O(0.1)$, the backward cascade cannot be seen as a simple 2D process but it is due to non-trivial three-dimensional coupling resulting in an accumulation of the energy not only toward the slow manifold $\bk_{_S}$ but also toward the small wavenumbers $k<k_f$. 

The backward cascade regime is extended in set A, comparing the spectra and the fluxes in this simulation (figures \ref{fig:spectra}(left) and \ref{fig:flux_slow_fast}(left) respectively), we can see that the three-dimensional coupling interaction measured in the fluxes are more important in the range, $k\lesssim k_f$, corresponding to the range where the energy spectrum has a scaling slope close to $k^{-3}$, far from the $k^{-5/3}$ typical of the 2D inverse cascade.
On the other hand, the forward cascade, developed in set B for $k>k_f$, is clearly carried 
by 3D interactions, as observed also in \cite{mininni2009scale}, with $\Pi_{_{S \leftrightarrows S}}(k)$ going quickly to zero at $k>k_f$. The forward flux is led by $\Pi_{_{F \leftrightarrows F}}(k)$, hence by the interactions inside the fast manifold plus a sub-leading contribution coming from the coupling between 
slow and fast manifolds, which in any case is a 3D effect. 

Recalling the discussion in Sec.~\ref{sec:slow-fast-dec}, the energy flux coupling slow and fast modes is 
composed of three different classes of triads (see Eq.~\eqref{eq:fluxfastslow}). Furthermore, one of them is a well defined energy flux (first term in the RHS Eq.~\eqref{eq:fluxfastslow}) while the remaining two are not conserving energy separately; only their sum can be seen as a conservative energy flux on the total volume. In this work we have analyzed the importance of these three different subsets separately. The first results we found is that the flux term, (first term in Eq.~\eqref{eq:fluxfastslow}) is very small, i.e. almost always zero, hence $\Pi_{_{F \leftrightarrows S}}(k)$ in Eq.~\eqref{eq:fluxfastslow} can be rewritten as;

\begin{equation}
\nonumber
\Pi_{_{F \leftrightarrows S}}(k) \approx \underbrace{ - \sum_{\substack{\bk \in W \\ |\bk|\le k}}  ik_j \hat{u}^*_i(\bk)\sum_{\substack{\bp\in V \\ \bq\in W}}  \hat{u}_i(\bp)\hat{u}_j(\bq) \delta(\bp+\bq -\bk) }_{\Pi_{_{F \leftrightarrows S}}^a(k)} \,\,
\underbrace{- \sum_{\substack{\bk \in V \\ |\bk|\le k}}  ik_j \hat{u}^*_i(\bk)\sum_{\substack{\bp\in W \\ \bq\in W}}  \hat{u}_i(\bp)\hat{u}_j(\bq) \delta(\bp+\bq -\bk)}_{\Pi_{_{F \leftrightarrows S}}^b(k)} \,.
\end{equation}
where $\Pi_{_{F \leftrightarrows S}}^a(k)$ comes from the Fourier transform of the mixed nonlinear term for the energy evolution equation inside the slow field in Eq.~\eqref{eq:slow-energy}, while $\Pi_{_{F \leftrightarrows S}}^b(k)$ comes from the evolution equation for the energy contained in the fast field, see Eq.~\eqref{eq:fast-energy}.

\begin{figure*}
\includegraphics[scale=0.59]{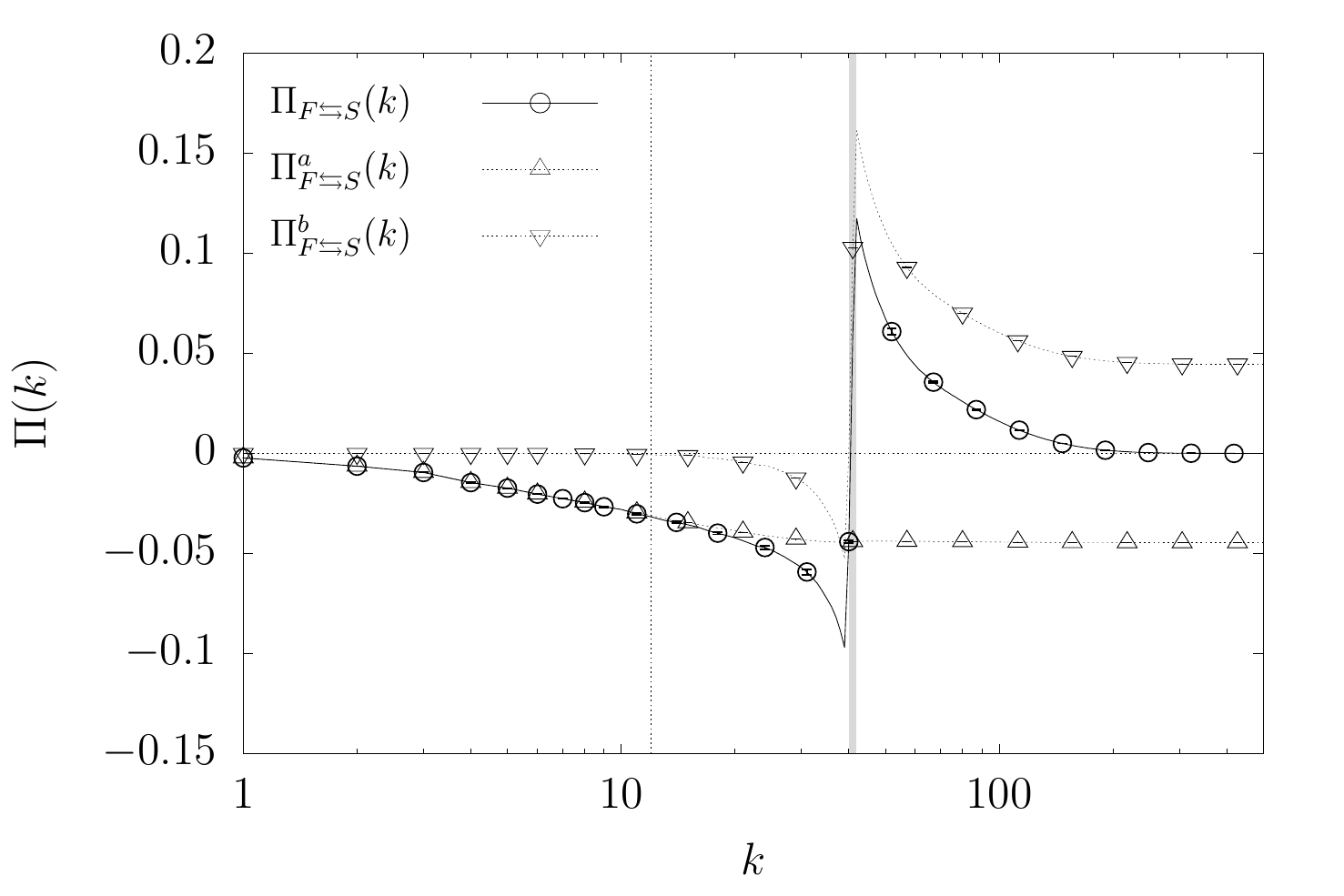}
\includegraphics[scale=0.59]{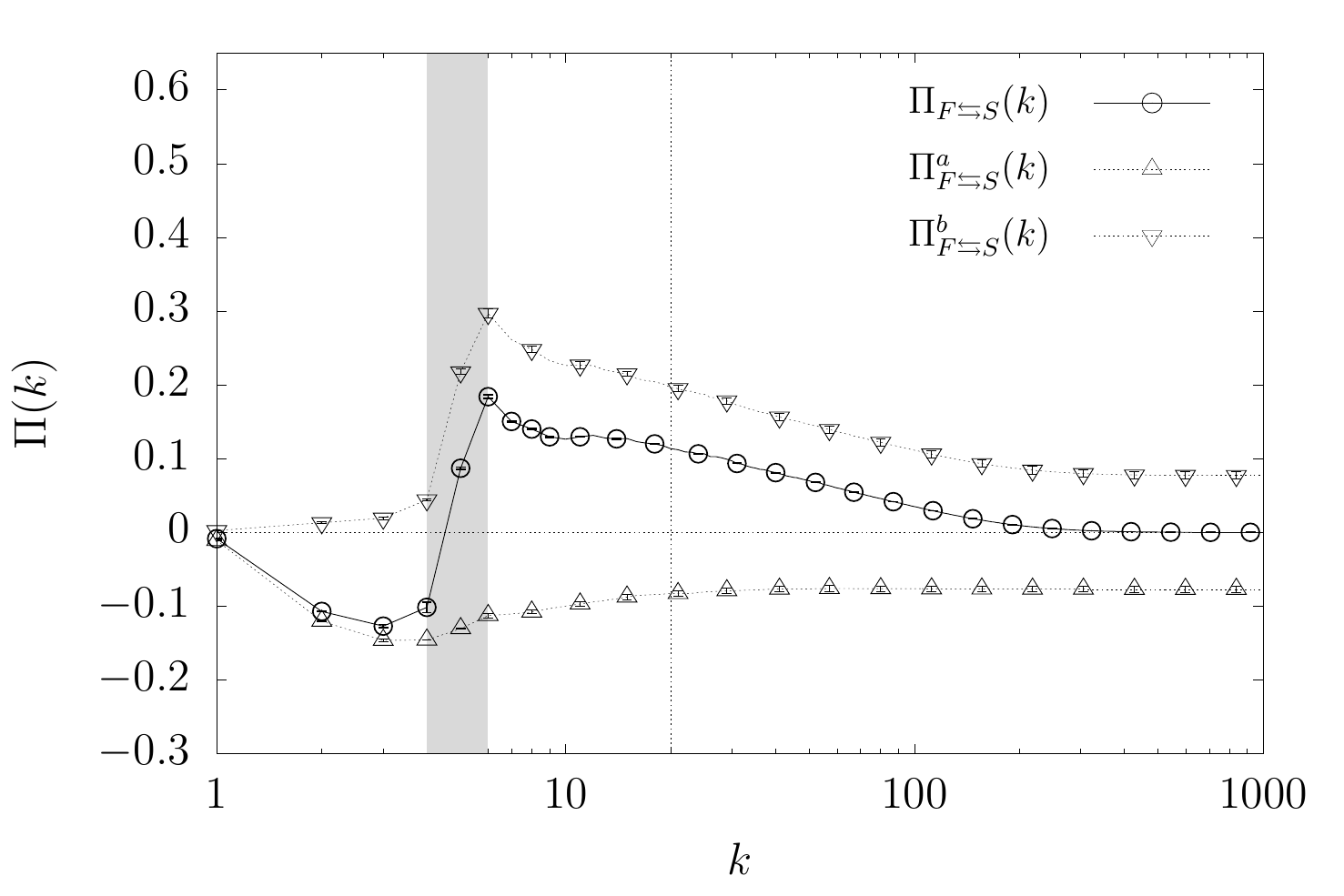}

\caption{Energy flux due to coupling slow-fast interactions, $\Pi_{_{F \leftrightarrows S}}(k)$ (empty circles) decomposed on its two different contributions, the one coming from the evolution equations of the slow modes (upwards triangles) and the one from the evolution equation of the fast modes (downwards triangles). (Left panel) Data from simulation A, (right panel) data from simulation B. \rep{The dashed vertical lines represent the cutoff scale, $k_c$, used in the analysis of the SGS energy transfer (see Sec.~\ref{sec:sgs_transfer})}.}
\label{fig:flux_coupl}
\end{figure*}

In Fig.~\ref{fig:flux_coupl} we show separately, $\Pi_{_{F \leftrightarrows S}}^a(k)$ and $\Pi_{_{F \leftrightarrows S}}^b(k)$ compared with the total coupling term shown in Fig.~\ref{fig:flux_slow_fast}. As already said the two terms are not separately conserving energy and they do not go to zero for $|\bk| \rightarrow \infty$. However it is interesting to study their signs. The backward regime of both simulations gives $\Pi_{_{F \leftrightarrows S}}^a(k)<0$ and $\Pi_{_{F \leftrightarrows S}}^b(k) \approx 0$, meaning that the energy is going backward into the slow manifold. On the other hand, in the forward cascade $k>k_f$, $\Pi_{_{F \leftrightarrows S}}^b(k)$ becomes positive meaning that there is net transfer of energy form the forcing to the small wavenumbers inside the fast modes due to coupling interactions.

\subsection{Helical mode analysis}
\begin{figure*}
\includegraphics[scale=0.59]{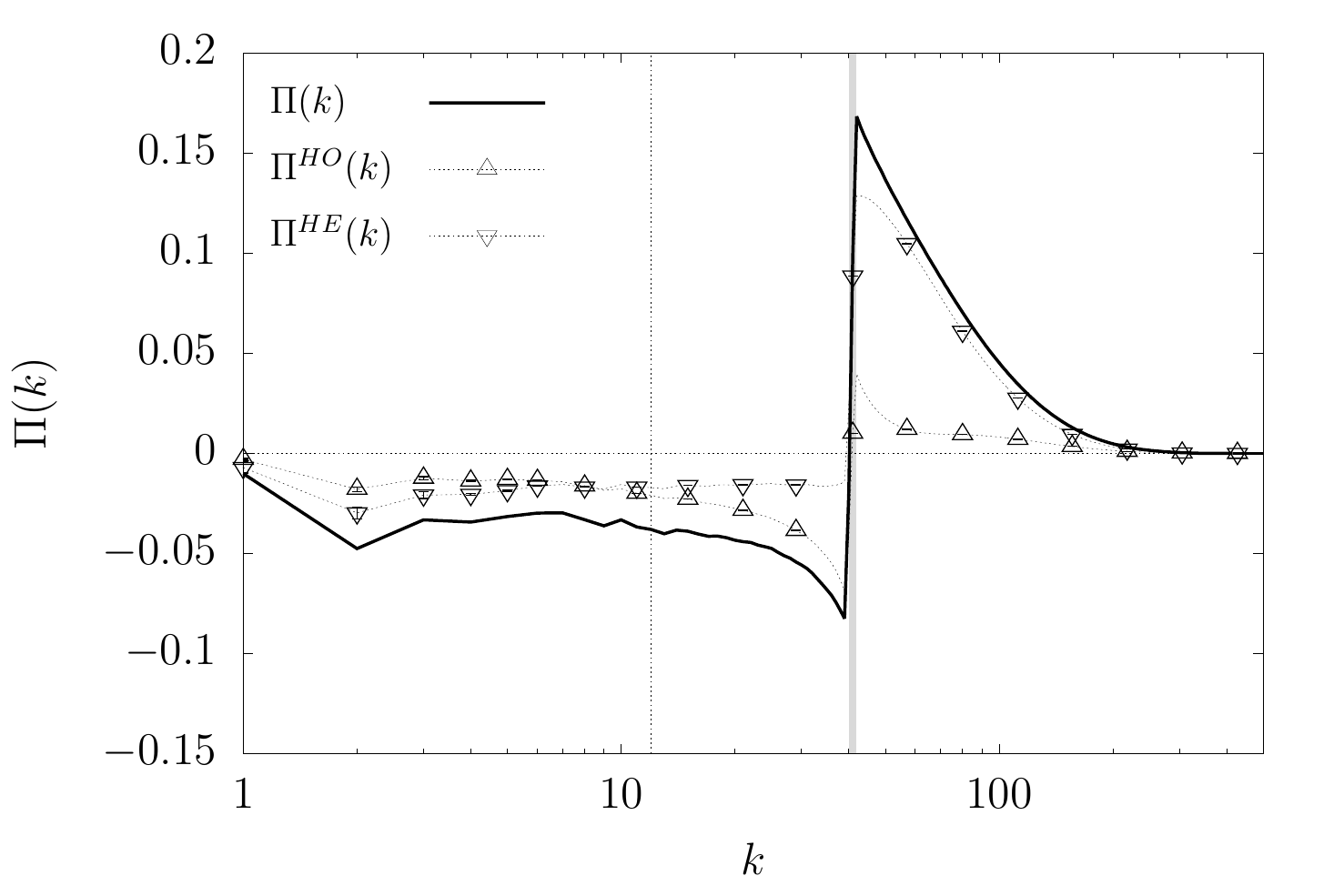}
\includegraphics[scale=0.59]{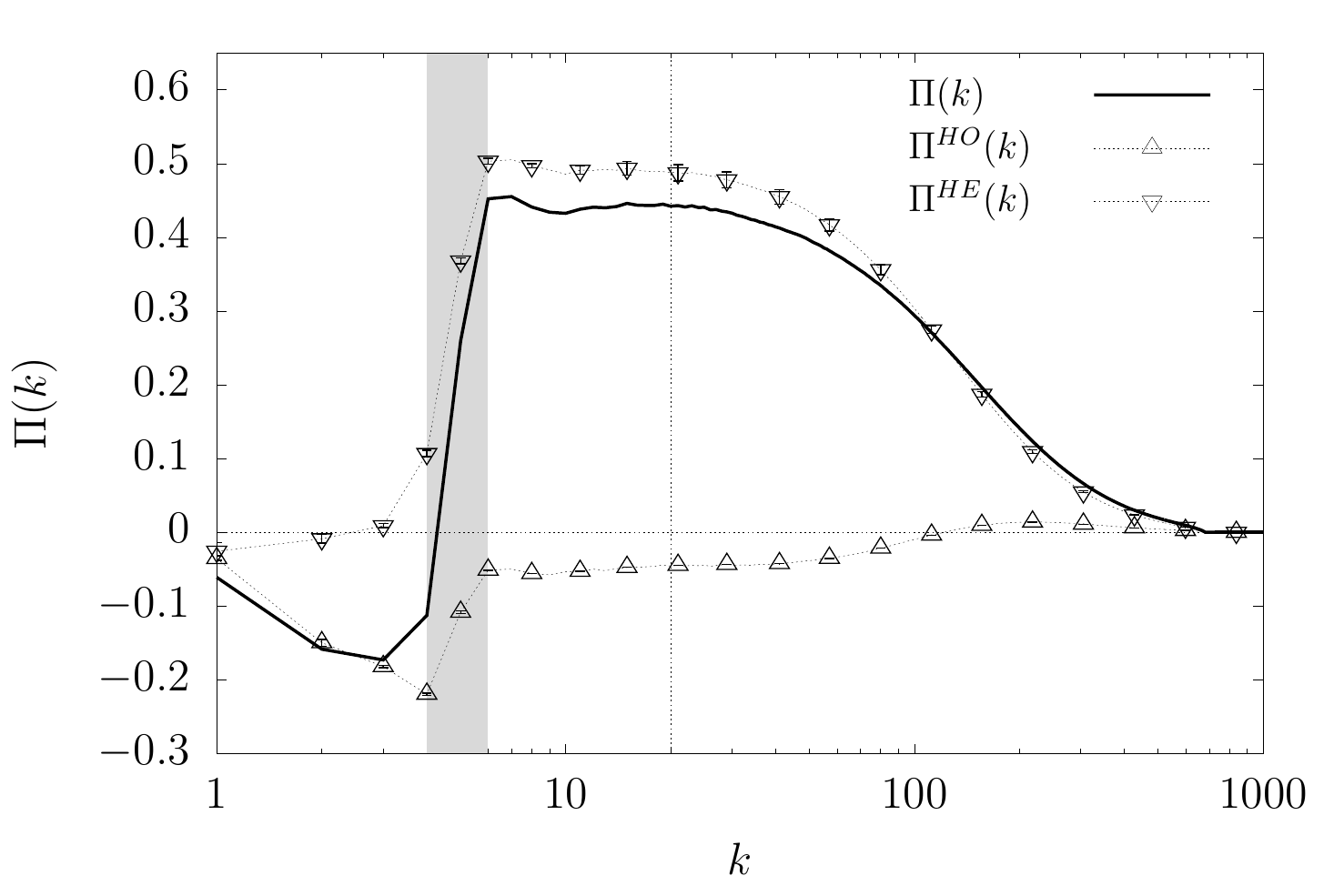}
\caption{Total energy flux $\Pi(k)$ (solid line) and fluxes decomposed into the homochiral, $\Pi^{HO}(k)$ (upward triangles), and heterochiral $\Pi^{HE}(k)$ (downwards triangles) bases. (Left panel) Data from simulation A, (right panel) data from simulation B. \rep{The dashed vertical lines represent the cutoff scale, $k_c$, used in the analysis of the SGS energy transfer (see Sec.~\ref{sec:sgs_transfer})}.}
\label{fig:flux_helicity}
\end{figure*}

To conclude the analysis of the energy flux in Fig.~\ref{fig:flux_helicity} we analyze the decomposition into homo- and heterochiral contributions. For set A, left panel, we see that the backward energy transfer, i.e. that at $k< k_f$, is mainly given 
by homochiral interactions in the range where the dynamics is dominated by three-dimensional coupling terms. This result is expected because in a three-dimensional domain only homochiral triads are known to produce an inverse cascade \cite{biferale2012inverse,biferale2013split}. At smaller wavenumbers instead, where we enter in the regime dominated by the interactions in the 2D slow plane, we have the same contribution coming from the homo- and heterochiral triads. This results is again in agreement with the previous results in \cite{biferale2017}, where it was shown that 
helicity cannot play a role in 2D dynamics because it always vanishes in that case.
In the right panel of the same figure, Fig.~\ref{fig:flux_helicity}, we can see that the forward cascade is generated by heterochiral triads.

\begin{table*}
\begin{center}
\begin{tabular}{|c c c c c c c c c|}
Set & $Ro$ \hspace{0.2cm} & {\color{white}kk}$k_{f}$ \hspace{0.3cm} &  {\color{white}kk}$ k^*$ \hspace{0.2cm} & $\Pi^{HO}/\Pi$ \hspace{0.2cm} & $\Pi^{HE}/\Pi$ \hspace{0.2cm} & $\Pi_{_{F \leftrightarrows F}}/\Pi$ \hspace{0.2cm} & $\Pi_{_{F \leftrightarrows S}}/\Pi$ \hspace{0.2cm}& $\Pi_{_{S \leftrightarrows S}}/\Pi$\\
\hline
 (A: Inverse) & 0.1 &[40:42] & 30 & 0.73 & 0.27 & 0.005 & 0.98 & 0.015 \\
 (C: Inverse)& 0.05 &[40:42] & 30 & 0.60 & 0.40 & 0.01 & 0.95 & 0.04\\
 (B: Split) & 0.1 &[4:6]  & 3  & 1.0 & 0.0 & 0.06 & 0.82 & 0.12\\
 (B: Split) & 0.1 &[4:6]  & 20 & -0.09 & 1.09 & 0.70 & 0.27 & 0.03\\
\end{tabular}
\end{center}
\caption{ \rep{The mean values of the energy flux decomposed on both the helical channels and the different slow-fast interactions normalized by the total flux are reported at $k=k^*$. The scales $k^*=30$ and $k^*=3$ respectively for the sets of simulations A, C, and B are chosen to estimate the fluxes in the backward regime, while the scale on set B is focused on the forward cascade regime.}}
\label{tbl:meanfluxes}
\end{table*}
\noindent 
\rep{We summarize the results coming from the Fourier space analysis of the energy flux in Table~\ref{tbl:meanfluxes}. Here we report the values of the mean flux decomposed on the different helical or slow-fast interactions normalized to the total energy flux at a scale $k=k^*$ in both directions of the energy cascade. For the inverse cascade we found that the homochiral interactions coupling the two slow-fast manifolds are dominant at scales close to the forcing. This means that in the parameter regimes used in all our simulations, realistic for geophysical flows, the energy relies on the homochiral interactions to be channeled up from the forcing scales to the slow manifold where it can then undergo an inverse cascade by 2D dynamics. However it is interesting to notice that at decreasing Rossby number (see Table~\ref{tbl:meanfluxes} simulation set C) the importance of the homochiral interactions is reduced, suggesting that in the limit of $Ro \rightarrow 0$ the dynamics may become purely 2D up to the forcing scales. Instead, in the forward cascade regime the total flux is found to be mainly produced inside the fast manifold through heterochiral interactions (see Table~\ref{tbl:meanfluxes} set B at $k^*=20$).}

\section{Sub-grid-scale energy transfer}
\label{sec:sgs_transfer}
To define a physical space energy transfer, we use a filtering approach common in large-eddy-simulations (LES) \cite{Pope00, Lesieur08, meneveauARFM}. We need to introduce a filtered velocity field and a filtered set of governing equations. In this way, we can define the scale separation in physical space and we can write the nonlinear structure of the coupling term among scales below and above the filter threshold. In this work, we use the `sharp spectral cutoff' filter in Fourier space \cite{Pope00} with a cutoff wavenumber $k_c$. 
The choice of a sharp spectral 
cutoff filter is convenient for two reasons. First, it is a projector which produces a clear scale separation. Second, there is an analytical equivalence between the mean energy transfer across the filter cutoff scale 
$\Delta = \pi/k_c$, 
and the Fourier space flux across the wavenumber $|\bk|=k_c$. The measure of the energy transfer in real space can then be achieved, following the coarse-graining procedure discussed  in \cite{buzzicotti2017effect} and outlined briefly here. 

For a filter kernel $G_{\Delta}(\bx)$, it is possible to obtain the filtered velocity field $\obu(\bx,t)$ by a convolution in real space between the filter kernel and the total velocity field $\bu(\bx,t)$,  
\begin{align}
\obu(\bx,t) &\equiv \int_\Omega  d\by \ G_\Delta(|\bx-\by|)\ \bu(\by,t) 
= \sum_{\bk \in \mathbb{Z}^3}  \hat G_\Delta(|\bk|)\ \bhu(\bk,t) e^{i \bk \bx} \ ,
\end{align}
with $\hat G_\Delta$ being the Fourier transform of $G_\Delta$, see Refs. \cite{Pope00, meneveauARFM}. To have access to the dynamical interaction existing between the coarse-grained and the sub-grid scales (SGS) velocity fields we need to apply the same filtering operation to the Navier-Stokes equations,

\be
\p_t \obu  + \nabla \cdot(\overline{\obu \otimes \obu}) + 2\Omega \times \obu
= -\nabla \oP -\nabla \cdot \otau(\bu,\bu)+ \nu \Delta \obu \ . 
\label{eq:Ples-rot}
\ee
where $\otau(\bu,\bu)$ is the SGS stress tensor, defined as;

\be
\label{eq:tau_ples}
 \otau_{ij}(\bu,\bu) =  \overline{u_iu_j} - \overline{\ou_i\ou_j} \ . 
\ee
and it contains the interactions between the scales above and below the filter threshold, $\Delta$.
Multiplying each term of Eq.~(\ref{eq:Ples-rot}) by the velocity field, we get an explicit formulation of the SGS energy transfer $\oPi$, 
the real contribution to the energy transfer across the filter scale.
Hence the kinetic energy balance in terms of $\oPi$ becomes:

\be
\label{eq:sg-eneP-leo}
\frac{1}{2}\p_t(\ou_i \ou_i) + \p_j A_{ij}   =  -\oPi -\Pileo \ ,
\ee
where $\partial_j A_{j} = \partial_j \ou_i ( \overline{\ou_i \ou_j} + \overline{p} \delta_{ij}+ \otau_{ij} - \frac{1}{2} \ou_i\ou_j)$ and $\Pileo = - \partial_j \ou_i \left ( \overline{\ou_i\ou_j}-\ou_i\ou_j \right )$. It is important to distinguish $\Pileo$ from the SGS energy transfer $\oPi$ because the former depends only on resolved-scale quantities and  does not contribute to the mean energy flux between sub grid and resolved scales (RS). In contrast, 
\be
\label{eq:SGS_Pi_ples}
 \oPi= -\partial_j \ov_i \,\, \otau_{ij}(\bv,\bv) = -\partial_j \ov_i \left ( \overline{v_iv_j} -  \overline{\ov_i\ov_j} \right ) \ , 
\ee
is the only flux which depends on both the SGS and the RS. 
Following the slow-fast decomposition introduced in Sec.~\ref{sec:slow-fast-dec}, we can introduce with the same procedure a real-space SGS energy transfer for each velocity component. In the following we will consider separately the contribution coming from the slow and fast manifolds and their coupling interactions. The three different contributions are defined as \cite{buzzicotti2017effect, carati2001}
\begin{align}
\label{eq:SGS_Pi_slow}
 \oPi_{_{S \leftrightarrows S}} =& -\partial_j \ou_{i}^{_S} \cdot \left ( \overline{u_{i}^{_S} u_{j}^{_S}} -  \overline{\ou_{i}^{_S} \ou_{j}^{_S}} \right ) \ , \\
\label{eq:SGS_Pi_fast}
 \oPi_{_{F \leftrightarrows F}} =& -\partial_j \ou_{i}^{_F} \cdot \left ( {\text P_{_F}} \left [\overline{u_{i}^{_F} u_{j}^{_F}} -  \overline{\ou_{i}^{_F} \ou_{j}^{_F}} \right ] \right ) \ , \\
 \oPi_{_{F \leftrightarrows S}} \approx -\partial_j \ou_{i}^{_F} \cdot ( \, &\overline{u_{i}^{_S} u_{j}^{_F}} -  \overline{\ou_{i}^{_S} \ou_{j}^{_F}} \, ) -\partial_j \ou_{i}^{_S} \cdot \left ( {\text P_{_S}} \left [ \overline{u_{i}^{_F} u_{j}^{_F}} -  \overline{\ou_{i}^{_F} \ou_{j}^{_F}} \right ] \, \right ) \ , 
\label{eq:SGS_Pi_coup}
\end{align}
Note that in Eq.~\eqref{eq:SGS_Pi_coup} we have used the approximation assuming $\langle \partial_j \ou_{i}^{_F} \cdot ( \, \overline{u_{i}^{_F} u_{j}^{_S}} - \overline{\ou_{i}^{_F} \ou_{j}^{_S}} \, ) \rangle \approx 0$, as discussed in Sec.~\ref{sect:Fspace}.

\section{Results: physical space analysis}

\begin{figure*}
\includegraphics[scale=0.59]{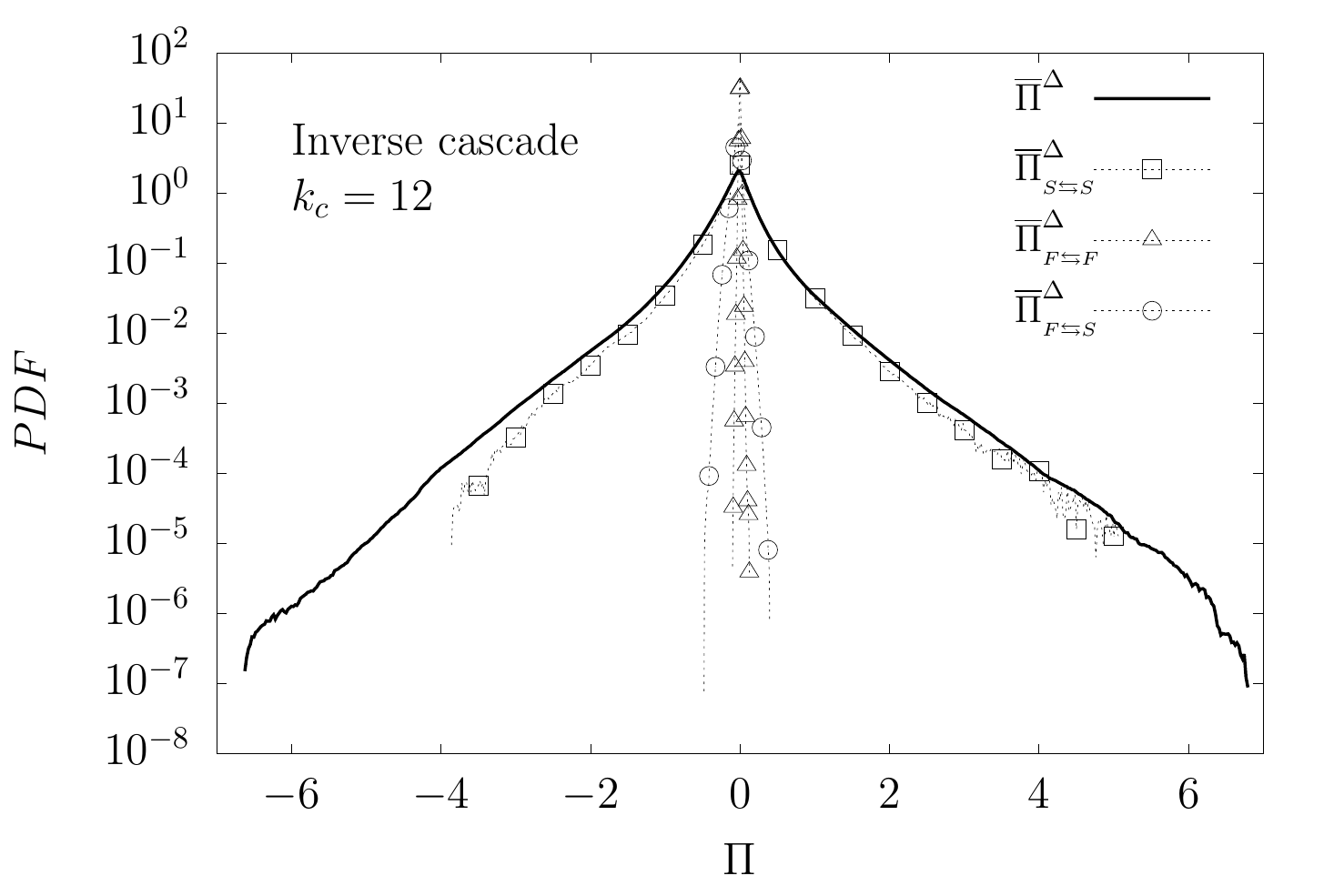}
\includegraphics[scale=0.59]{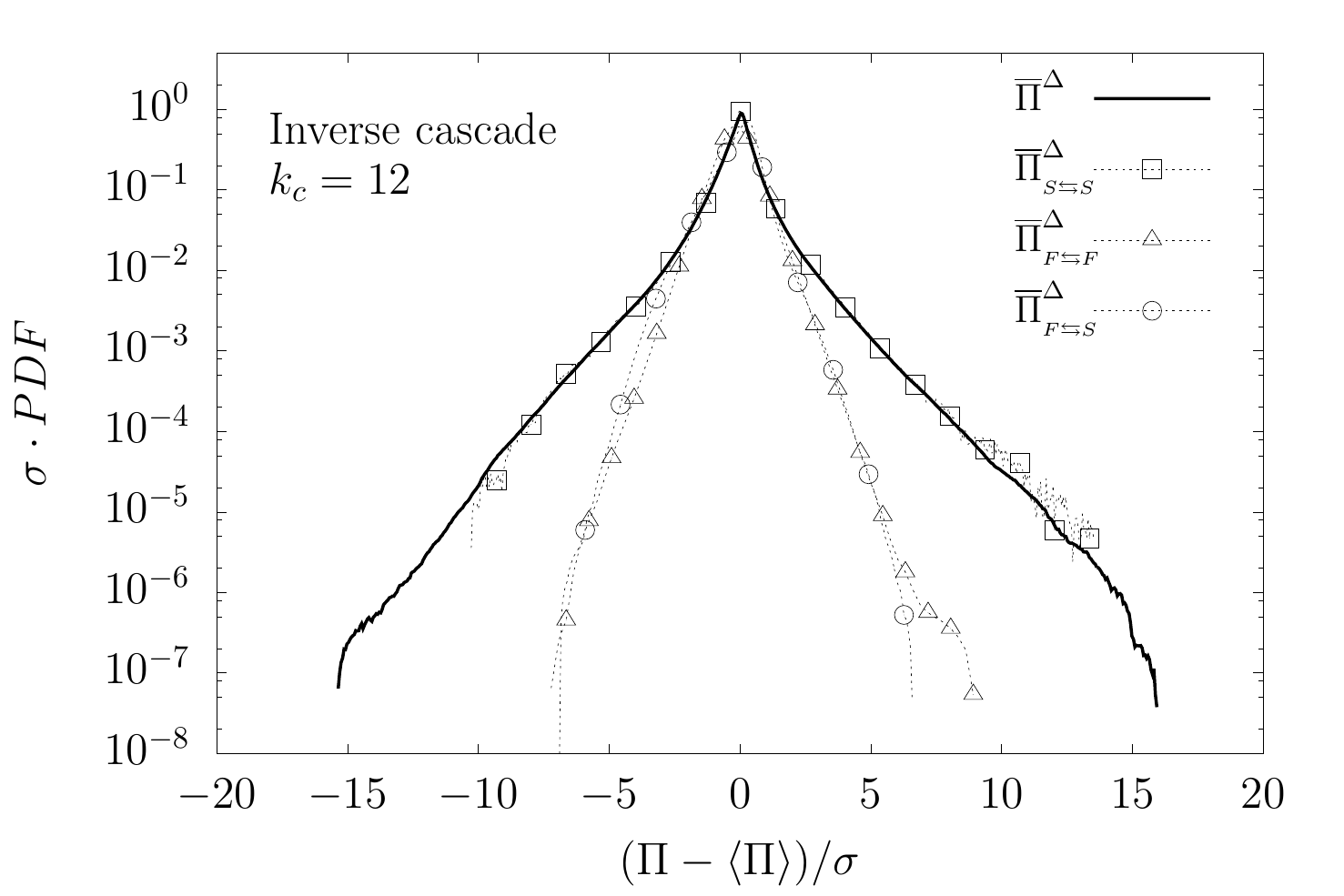}

\caption{(Left panel) PDFs of SGS energy transfer for the inverse cascade simulation, decomposed on the different slow-fast interactions; total interactions $\oPi$ (solid line), slow-slow interactions $\oPi_{_{S \leftrightarrows S}}$ (empty squares), fast-fast interactions $\oPi_{_{F \leftrightarrows F}}$ (empty triangles) and coupling slow-fast interactions $\oPi_{_{F \leftrightarrows S}}$ (empty circles). (Right panel) Standardized PDF to zero mean and $\sigma = 1$ for the same data. \rep{All the SGS energy transfer are measured with the cutoff at $k_c=12$, in the inverse cascade regime of simulation A. The values for the skewness ($S$) and the flatness ($F$) for the PDF of the total interactions ($\oPi$) are: $S= \langle x^3 \rangle/\langle x^2 \rangle^{3/2} \simeq -0.42$ and $F= \langle x^4 \rangle/\langle x^2 \rangle^2 \simeq 14$.}}
\label{fig:PDF_A}
\end{figure*}

\begin{figure*}
\hspace{-0.cm}
\includegraphics[scale=0.42]{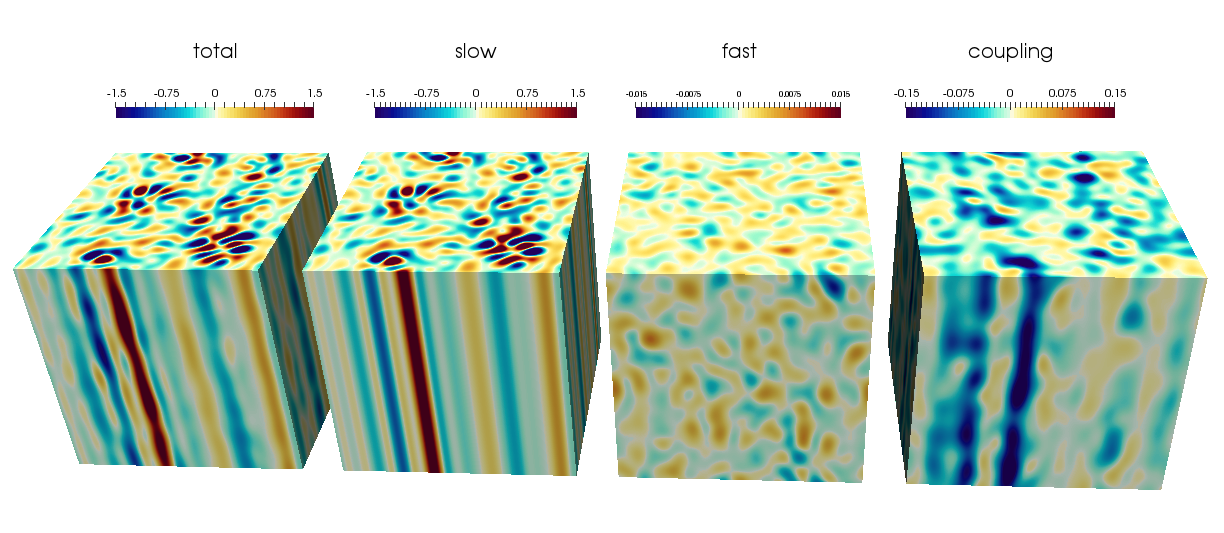}

\caption{Visualization of the total SGS energy transfer $\oPi$, across the filter scale corresponding to $k_c=12$, 
and its three terms coming from the slow-fast decomposition, namely; slow $\oPi_{_{S \leftrightarrows S}}$, fast $\oPi_{_{F \leftrightarrows F}}$ and coupling $\oPi_{_{F \leftrightarrows S}}$. 
The data corresponds to simulation A with the inverse cascade.}
\label{fig:4pi_A}
\end{figure*}

\begin{figure*}
\includegraphics[scale=0.59]{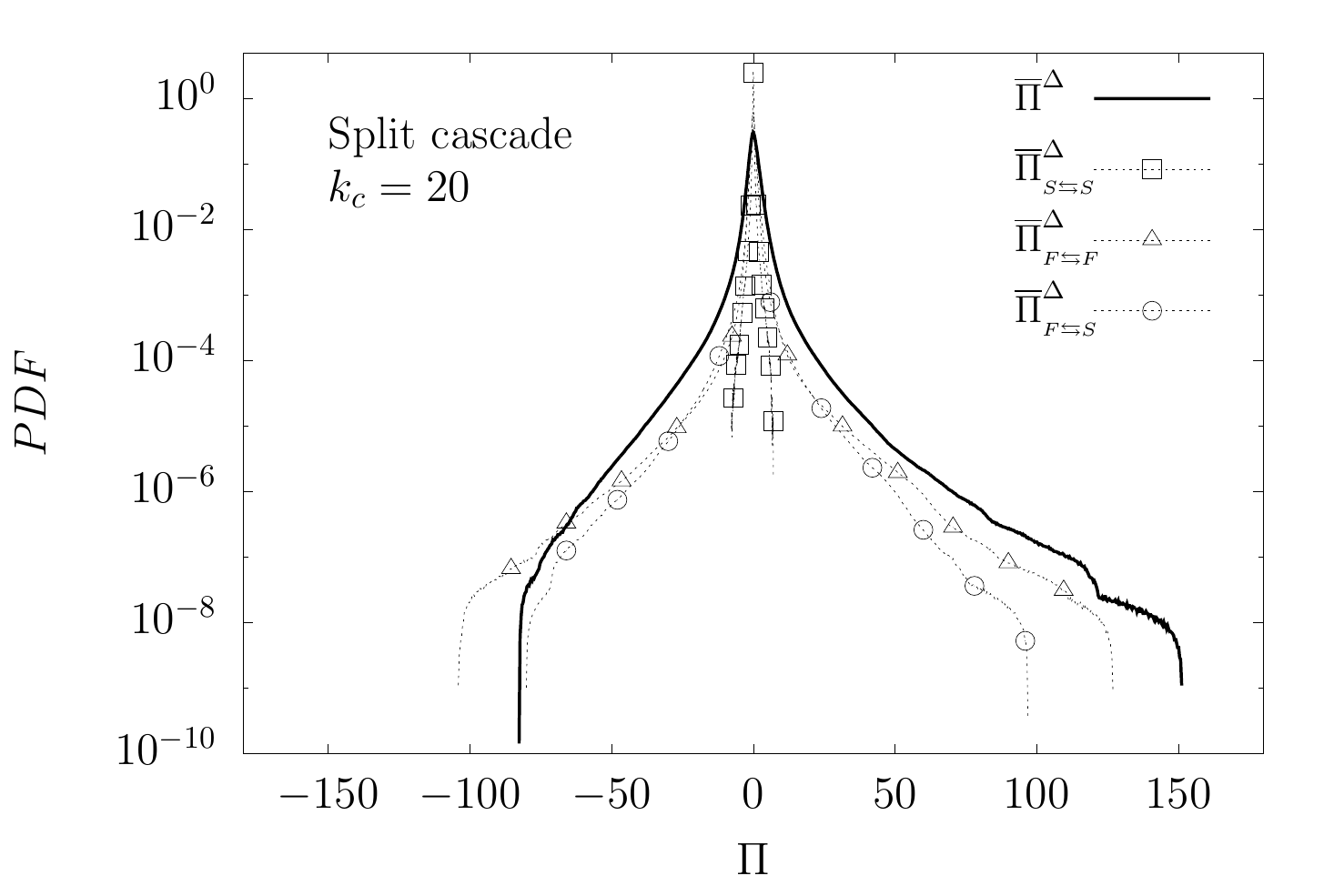}
\includegraphics[scale=0.59]{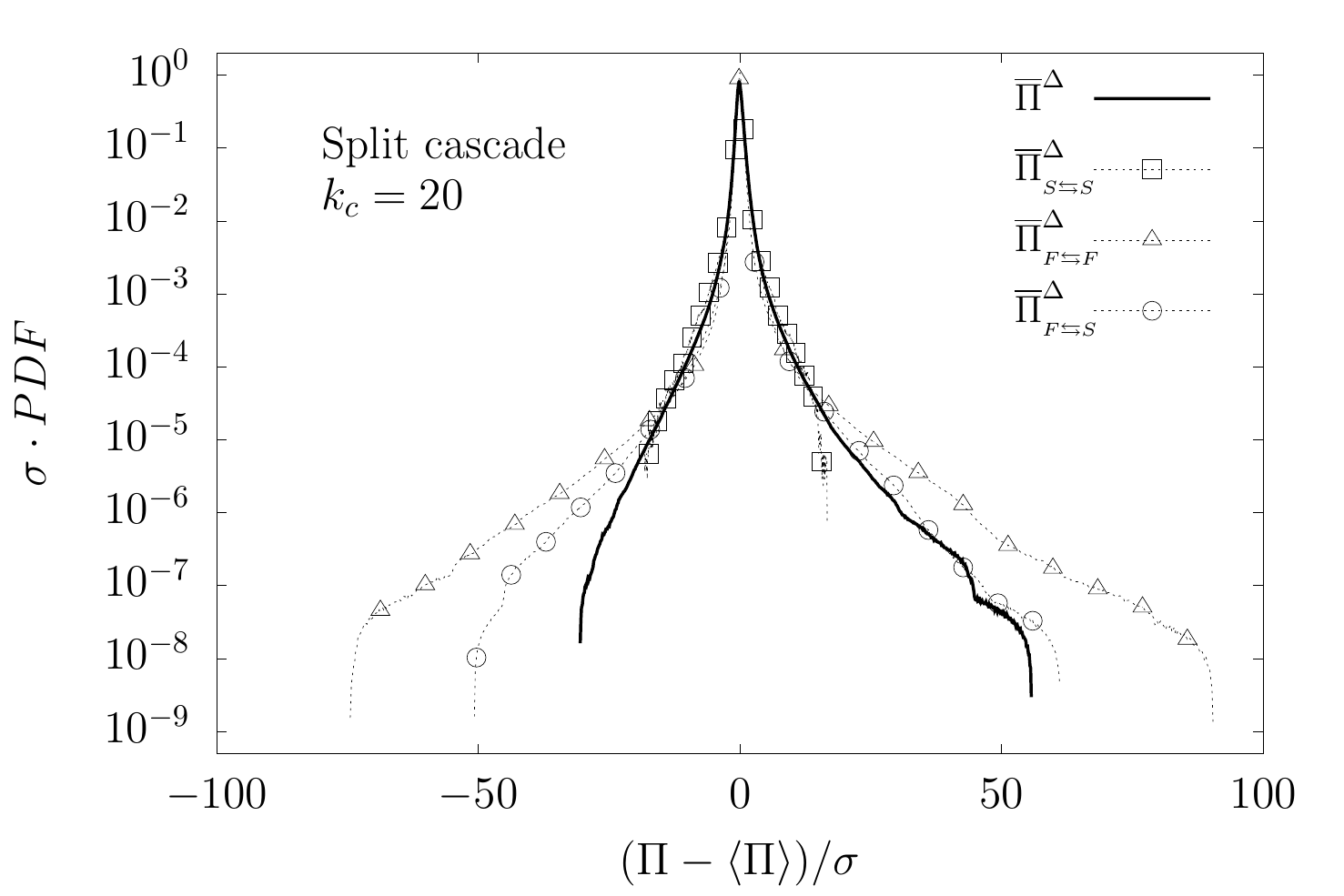}
\caption{(Left panel) PDFs of SGS energy transfer for the split cascade simulation, decomposed on the slow/fast different interactions; total interactions $\oPi$ (solid line), slow-slow interactions $\oPi_{_{S \leftrightarrows S}}$ (empty squares), fast-fast interactions $\oPi_{_{F \leftrightarrows F}}$ (empty triangles) and coupling slow-fast interactions $\oPi_{_{F \leftrightarrows S}}$ (empty circles). (Right panel) Standardized PDF to zero mean and $\sigma = 1$ for the same data. \rep{All the SGS energy transfer are measured with the cutoff at $k_c=20$, in the direct cascade regime of simulation B. The values for the skewness ($S$) and the flatness ($F$) for the PDF of the total interactions ($\oPi$) are: $S= \langle x^3 \rangle/\langle x^2 \rangle^{3/2} \simeq 1.5$ and $F= \langle x^4 \rangle/\langle x^2 \rangle^2 \simeq 68$.}}
\label{fig:PDF_B}
\end{figure*}

\begin{figure*}
\hspace{-0.cm}
\includegraphics[scale=0.46]{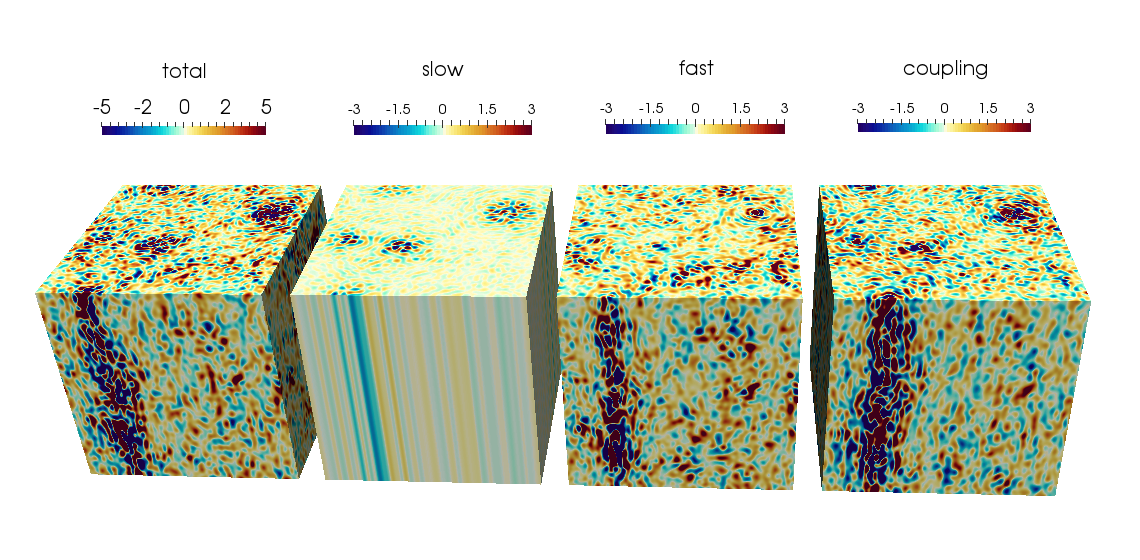}
\caption{Visualization of the total SGS energy transfer $\oPi$, across the filter scale corresponding to $k_c=20$, and its three 
contributions 
coming from the slow/fast decomposition, namely; slow $\oPi_{_{S \leftrightarrows S}}$, fast $\oPi_{_{F \leftrightarrows F}}$ and coupling $\oPi_{_{F \leftrightarrows S}}$. 
The data corresponds to simulation B with a split energy cascade.}
\label{fig:4pi_B}
\end{figure*}

\noindent 
In this section we present the physical space analyzes on the statistics of the SGS energy transfer after decomposing the interaction of the slow and fast manifolds. We present the probability distribution functions (PDFs) of $\oPi$ compared with its separate contributions, namely; $\oPi_{_{S \leftrightarrows S}}$, $\oPi_{_{F \leftrightarrows F}}$ and $\oPi_{_{F \leftrightarrows S}}$ see Eqs.~\eqref{eq:SGS_Pi_slow},~\eqref{eq:SGS_Pi_fast} and~\eqref{eq:SGS_Pi_coup}.
In the following we use a sharp projector filter to separate the resolved from the sub-grid scales. 
In particular for the analysis of the inverse cascade we consider set A with a cutoff at $k_c=12$, while in the forward case using data from simulation set B, we apply a cutoff at $k_c=20$.

In Fig.~\ref{fig:PDF_A} the PDF and the PDF normalized to their standard deviation measured on the dataset A are presented. We can see that fluctuations in the backward regime are dominated by slow-slow interactions. This result 
may appear obvious  
considering that the energy contained in the slow manifold at wavenumbers $|\bk|=12$ is an order of magnitude larger than the energy contained in the fast field, (see spectra in Fig.~\ref{fig:spectra}(left panel)) and considering that $\oPi_{_{S \leftrightarrows S}}$ is the only term defined as the product of three slow fields, (see Eq.~\eqref{eq:SGS_Pi_slow}), while all the other interactions couple 
at least two fast fields, see Eq.~\eqref{eq:SGS_Pi_fast}-\eqref{eq:SGS_Pi_coup}. However comparing the normalized PDFs, Fig.~\ref{fig:PDF_A}(right panel), we find that the different terms are all characterized by large and symmetric fluctuations up to 10-15 times the standard deviation, which are able to locally mask the information about the mean energy transfer obtained after an integration over the volume.

In Fig.~\ref{fig:4pi_A} the visualizations of 
$\oPi$ and its different components 
at a fixed time are shown to give an idea of the local distribution of extreme events of energy transfer measured in the PDF. 
From this figure it is worth noticing that the coupling term is mostly covered by the light blue color, indicating a backward energy transfer with an intensity of the order of $0.075$ comparable with the mean energy transfer at $k=12$ shown in Fig.~\ref{fig:flux_slow_fast}, left panel. 
This observation would suggest that in this configuration the coupling term is the best representative of the mean properties of the energy transfer.

In Fig.~\ref{fig:PDF_B} 
the same type of PDFs are presented for the forward cascade regime using dataset B with a cutoff at $k=20$. In this regime, we observe fewer Gaussian PDFs with fatter tails compared to the backward case, suggesting that the forward regime has a better energy flux which is visible not only on average but also on the local fluctuations. Looking at the tails, however, we see again fluctuations $10^3$ times larger than the mean value, (which is $O(0.1)$ see Fig.~\ref{fig:flux_coupl}). Here, in contrast, the less energetic modes are the slow ones, and hence the fluctuations due to slow-slow interactions are completely subleading compared to $\oPi_{_{F \leftrightarrows F}}$. 
The standard PDFs show even larger tails in the forward case, where values around 100 times larger than the standard deviation have been measured.
\rep{The skewness toward the right tail in the PDF of $\oPi$, tells us that extreme events are more probable to happen toward the direction of the mean energy transfer, which is on average positive at the cutoff scale $k_c=20$ of dataset B.}

In Fig.~\ref{fig:4pi_B} the visualizations at fixed time for the different SGS energy transfers are presented, note that the extreme fluctuations are produced inside the 
Taylor columns for all types of interactions.

\subsection{Q-criterion}

In the following, we measure the role and the relative importance of the regions dominated by strain compared to the regions dominated by vorticity (i.e. the regions in physical space occupied by the big columnar vortices), on the local and mean energy transfer across scales. In order to distinguish these areas in the physical domain we use the `Q-criterion' first introduced in Ref. \cite{hunt1988eddies} and which has been used extensively since then (e.g. Ref. \cite{burger2012vortices}). The criterion is based on the scalar field
\be
Q(\bx)= \frac{1}{8} \left ( |\nabla \times \bu|^2  - |\nabla \bu + (\nabla \bu)^T|^2 \right ),
\ee
which is the velocity gradient's second invariant.  It allows the identification of regions dominated by vorticity where $Q>0$, while  regions dominated by strain are characterized by a negative value of $Q$.

\begin{figure*}
\includegraphics[scale=0.59]{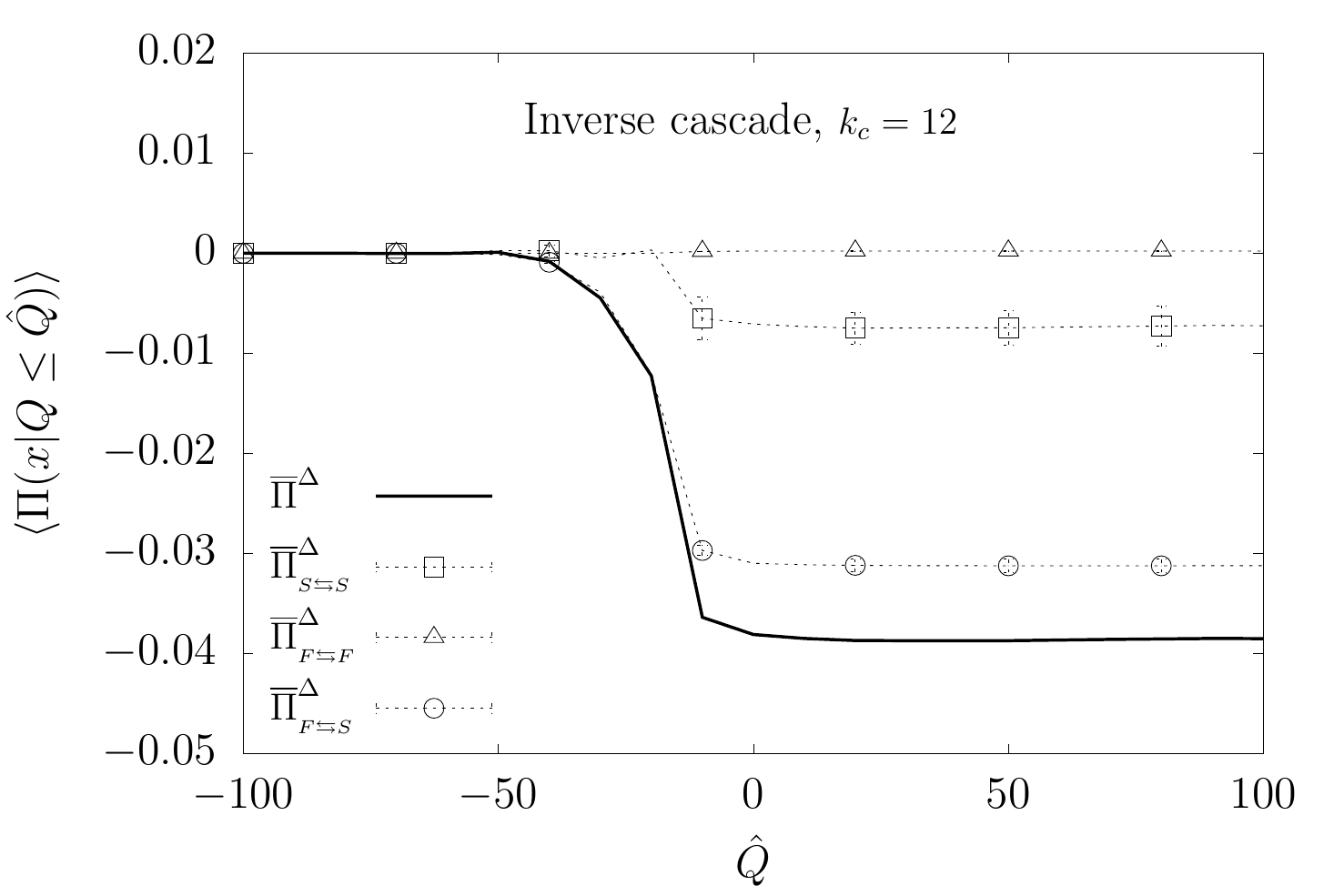}
\includegraphics[scale=0.59]{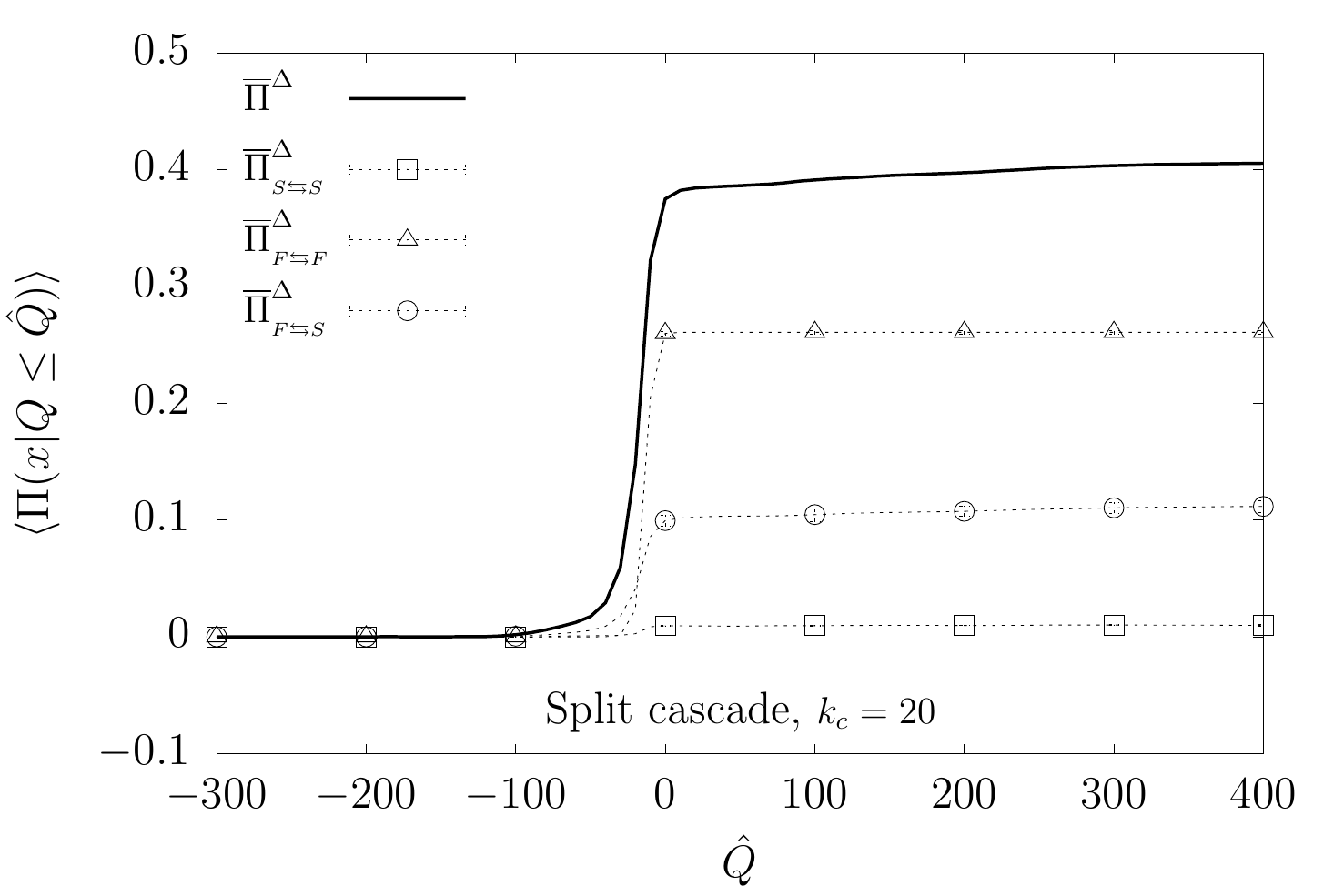}
\caption{Mean value of SGS energy transfer, $\langle \oPi \rangle$, conditioned on the volume's regions where $Q \le \hat{Q}$ (solid line). The same analysis is done considering only slow-slow interactions $\langle\oPi_{_{S \leftrightarrows S}} \rangle$ (empty squares), fast-fast interactions $\langle\oPi_{_{F \leftrightarrows F}} \rangle$ (empty triangles) and coupling slow-fast interactions $\langle\oPi_{_{F \leftrightarrows S}} \rangle$ (empty circles). (Left panel) Data from dataset A with a cutoff at $k_c=12$, and (right panel) data from dataset B with a cutoff at $k_c=20$. In both cases the field $Q$ is measured from a filtered velocity field with energy up to $k=7$.}
\label{fig:Run_Int}
\end{figure*}

\begin{figure*}
\includegraphics[scale=0.59]{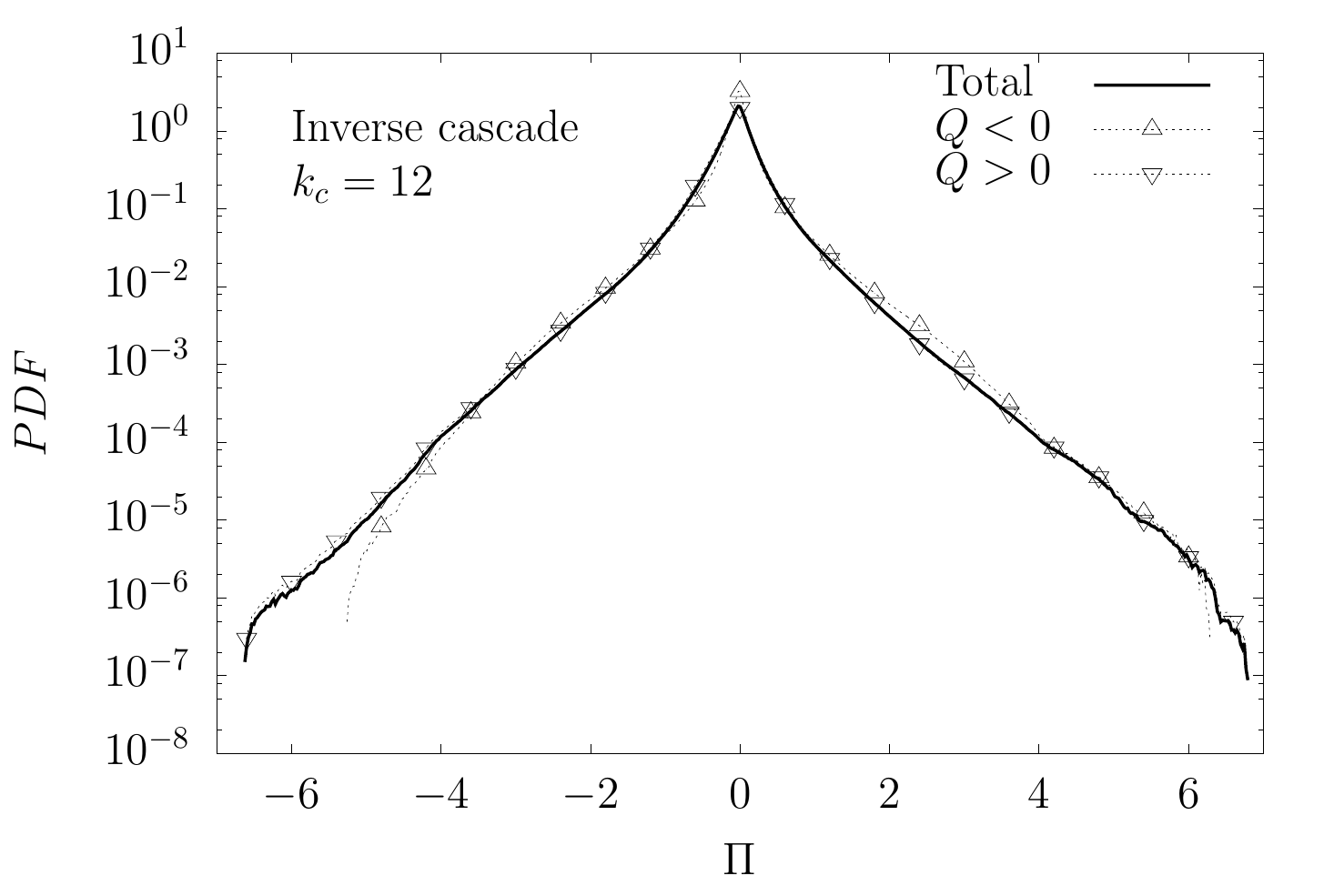}
\includegraphics[scale=0.59]{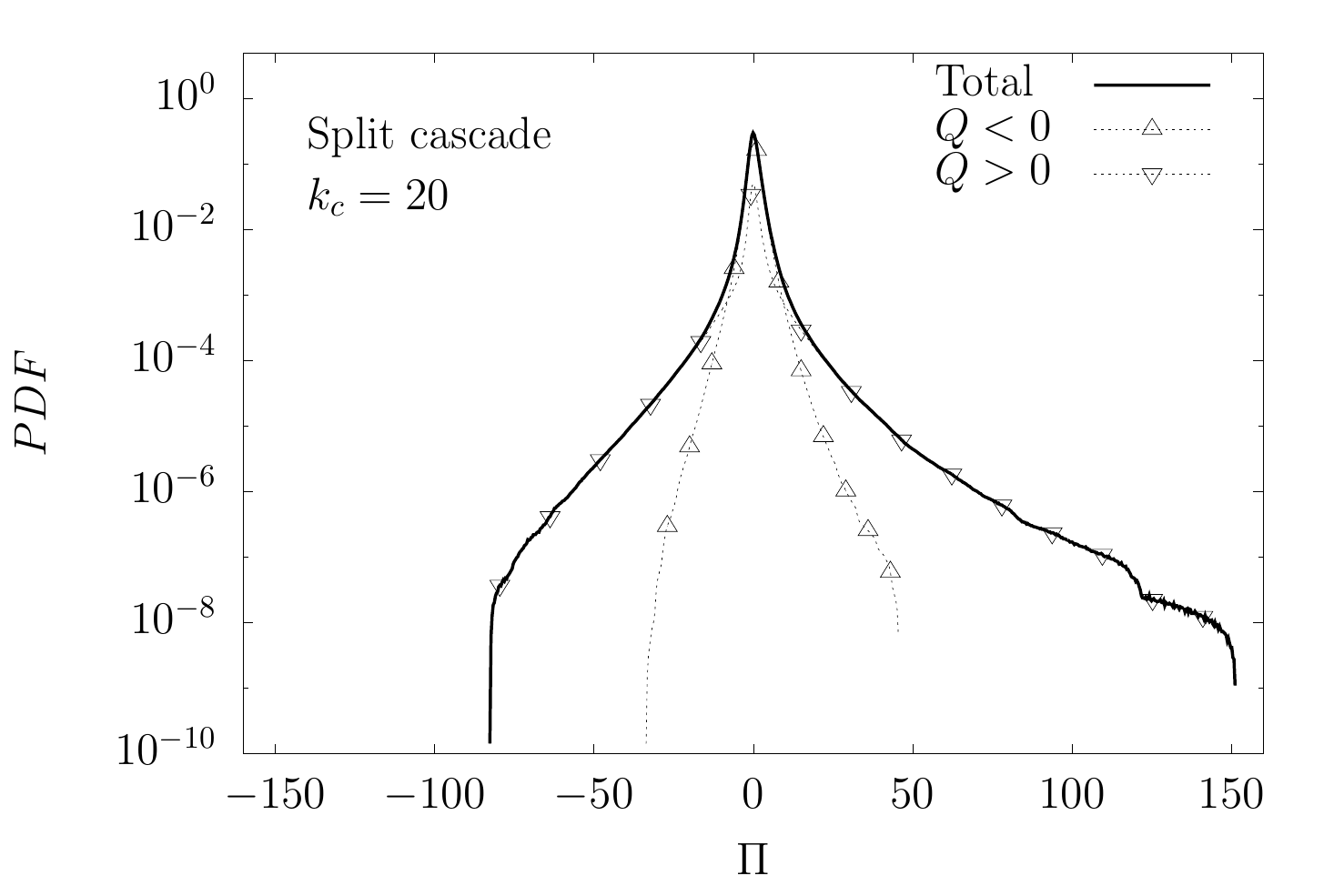}
\caption{PDFs of SGS energy transfer $\oPi$ (solid line) and the same PDF measured only on regions of space dominated by vorticity, $Q>0$, (downwards triangles) and on regions dominated by strain, $Q<0$, (upward triangles). (Left panel) Data from dataset A with a cutoff at $k_c=12$, and (right panel) data from dataset B with a cutoff at $k_c=20$. In both cases the field $Q$ is measured from a filtered velocity field with energy up to $k=7$.}
\label{fig:Pi_cond_Q}
\end{figure*}

To have a clearer identification of the large-scale vortices without contamination from small-scale structures, we calculate $Q$ using the low-pass filtered velocity field. To filter out the small scales we use a sharp filter with a cutoff at $k=7$. Using the resultant $Q(\bx)$ field, we can measure its spatial correlation with SGS flux. In particular, we measure the mean $\langle \oPi \rangle$ conditioned on different levels of $Q(\bx)$:
\be
\langle \oPi(\bx | Q \le \hat{Q}) \rangle = \frac{1}{V}\int_{V} \oPi(\bx | Q \le \hat{Q}) d\bx .
\ee
When $\hat{Q}$ is equal to the maximum $Q$, we recover the mean value over the entire domain, which equals the Fourier space energy flux across the filter wavenumber 
$k = \pi/\Delta$.
The measurements of $\langle \oPi(\bx | Q \le \hat{Q}) \rangle$ obtained from the two datasets are presented in Fig.~\ref{fig:Run_Int}. 
\rep{From Fig.~\ref{fig:Run_Int} we can see that both in the inverse and in the split cascade regimes, the mean energy properties are given by the strain regions with $Q(x)<0$. In Fig.~\ref{fig:Pi_cond_Q}, we further decompose the PDF of SGS energy transfer for the two regimes (inverse or forward) conditioning on being in a strain $Q<0$ or rotating region ($Q>0$).   Remarkably enough,  we observe that the fluctuations of the SGS energy transfer for the forward regime  (Fig.~\ref{fig:Pi_cond_Q}, right panel) show a different trend, with larger fluctuations when inside the strong vortical regions ($Q>0$). Concerning the inverse cascade regime (Fig.~\ref{fig:Pi_cond_Q}, left panel) we do not measure any difference inside rotating or strain regions. Both results support the indication that the intense fluctuations of the SGS energy transfer are not locally correlated with the same topological structures that contribute to the mean component.}


\section{Conclusions}

In this work, we assessed the scale-by-scale energy transfer in a fully developed rotating turbulent flow by means of state-of-the-art numerical simulations. We conducted two sets of simulations with different forcing scales and angular rotation rates such that in one case the energy flux is upscale while in the other case there is a split cascade going simultaneously upscale and downscale from that of the forcing. We measured the Fourier space fluxes in order to quantify the effects of rotation on its mean properties. We performed a physical space investigations in order to distinguish the dynamical role of regions dominated by columnar vortices from regions dominated by strain.

From the Fourier space analysis, we identified the structure and the geometry of interactions leading to the mean inverse energy transfer. Two possible scenarios are known to provide such behavior; the dynamics follows either (i) two-dimensional channels given by the triads constrained to live on the plane perpendicular to the rotation axis, $\bk=(k_x,k_y,k_z=0)$ (``slow manifold''), or (ii) fully three-dimensional interactions coupling triads with definite chirality (``homochiral triads''). 
To evaluate the relative weight of the different contributions to the total flux, we projected the flow field on (i) the slow-fast manifolds, and on (ii)  the homo- and the heterochiral subsets. 

We found that the accumulation of energy on the 2D slow manifold is produced by a three-dimensional mechanism involving triads connecting the slow with the fast manifolds. In particular, this behavior is more pronounced  close to the forcing-scales where it is supported by an imbalance between homo and heterochiral interactions. Moving toward the smallest wave numbers, where the energy is almost fully contained in the slow manifold, the two-dimensional interactions become the only ones responsible for the backward transfer. This observation is supported by the fact that the contributions on the inverse flux coming from homo- and heterochiral triads become indistinguishable at those wave numbers, as expected in the 2D dynamics. In contrast, we found that the direct cascade is always driven by triadic heterochiral interactions living inside the 3D fast manifold.

From the physical space analysis of the energy transfer, we quantified the relative influence of intense regions dominated by strong coherent vortices or by high values of the strain. In the definition of the physical space energy transfer we followed the approach used in LES based on the introduction of a filtered velocity field. For the spatially local SGS transfer, we distinguished the strain regions from the vortical ones using the Q-criterion (see Refs. \cite{hunt1988eddies,burger2012vortices}). We observed that the fluctuations overwhelm the mean value of the transfer by several orders of magnitude. 

Upon performing the spatial decomposition and a conditional average of the SGS energy transfer on regions dominated by strain or by  vorticity, we observed that the mean transfer can be completely reconstructed by the strain regions only. The situation changes for the conditioned PDF of the SGS energy transfer, where we find that the extreme events are dominated by the more energetic vortical regions. The latter was more evident in the forward cascade regime where the PDF appears to be skewed toward the right tail in accordance with the cascade direction. 
On the other hand, in the inverse cascade, the intensity of the fluctuations measured inside or outside the vortices is comparable, suggesting that the inverse energy flux is less efficient and closer to equilibrium compared to the forward cascade. \rep{This observation is in a qualitative agreement with the absence of intermittency in the inverse cascade regime.}
In conclusion, we have found that the inverse energy transfer in rotating turbulent flows is the result of combined 3D (homochiral) and 2D effects and that the exact decoupling among slow and fast manifolds expected in the limit of very small Rossby number is not observed at the rotation rates investigated here.
\rep{A possible extension of this work consists in the assessment of the energy transfer sensitivity on the chirality of the external forcing. A similar analysis in this direction has been performed in the case of a 2D3C flow, where the inverse energy cascade has shown to be strongly dependent on the level of helicity injected by the forcing~\cite{linkmann2018nonuniversal}. The analysis of the role of the forcing chirality in the context of turbulence under rotation is left for future work.}

\section*{Acknowledgments}
The research leading to these results has received funding from the European
Union's Seventh Framework Programme (FP7/2007-2013) under Grant Agreement No.
339032.
H.A. was also supported through NSF Grant No. OCE-1259794, DOE Grants No. DE-SC0014318
and No. \proof{DE-NA0001944}, and the LANL LDRD program through Project
No. 20150568ER.

\bibliographystyle{unsrt}
\bibliography{refs}

\end{document}